\documentclass{aa}  
\usepackage{natbib}
\usepackage{graphicx}
\usepackage{txfonts}
\usepackage{xcolor}
\usepackage{placeins} 

\usepackage{caption}
\usepackage{wrapfig}
\usepackage{siunitx}
\usepackage{amsmath}

\usepackage{natbib,twoopt}
\definecolor{links}{rgb}{0, 0, 255}
\usepackage[colorlinks=true,allcolors=links]{hyperref} 
\bibpunct{(}{)}{;}{a}{}{,}   

\newcommand{\simba}{\textsc{Simba}}

\newcommand{\HI}{\ion{H}{i}}

\newcommand{\hmpc}{h^{-1}{\rm Mpc}}

\newcommand{\kms}{\;{\rm km}\,{\rm s}^{-1}}

\newcommand{\msun}{{M}_{\odot}}

\begin{document}

\title{Investigating the residuals in the $M_\bullet - M_*$ relation using the \simba ~cosmological simulation} 
\titlerunning{Residuals in the $M_{\bullet}\ - \ M_*$ relation}

\author{
Wenlin Ma\inst{\ref{inst1},\ref{inst2}\fnmsep\thanks{Corresponding Author; mawenlin@shao.ac.cn}}
\and Weiguang Cui \inst{\ref{inst3},\ref{inst4},\ref{inst5}\fnmsep\thanks{Talento-CM fellow; weiguang.cui@uam.es}}
\and Romeel Dav\'e \inst{\ref{inst5}}
\and Daniel Angl\'es-Alc\'azar \inst{\ref{inst6}}
\and Hong Guo \inst{\ref{inst1}}
}

\institute{
Shanghai Astronomical Observatory, Chinese Academy of Sciences, Shanghai 200030, China. \label{inst1}
\and the University of Chinese Academy of Sciences, Beijing 100049, China. \label{inst2} 
\and Departamento de F\'isica Te\'orica, M-8, Universidad Aut\'onoma de Madrid, Cantoblanco, E-28049, Madrid, Spain. \label{inst3}
\and Centro de Investigaci\'on Avanzada en F\'isica Fundamental (CIAFF), Universidad Aut\'onoma de Madrid, Cantoblanco, E-28049 Madrid, Spain. \label{inst4}
\and Institute for Astronomy, University of Edinburgh, Royal Observatory, Edinburgh EH9 3HJ, United Kingdom. \label{inst5}
\and Department of Physics, University of Connecticut, 196 Auditorium Road, U-3046, Storrs, CT, 06269, USA\label{inst6}
}

\abstract
  {We study the scaling relation between the black hole and stellar mass ($M_\bullet-M_*$), diagnosing the residual $\Delta \log(M_\bullet/\msun)$ ($\Delta$) in this relation to understand the coevolution of the galaxy and black hole (BH) in the cosmological hydrodynamic simulation \simba. We showed that \simba ~can reproduce the observed $M_\bullet-M_*$ relation well with little difference between central and satellite galaxies. By using the median value to determine the residuals, we found that the residual is correlated with galaxy cold gas content, star formation rate, colour and black hole accretion properties. Both torque and Bondi models implemented in \simba, contribute to this residual, with torque accretion playing a major role at high redshift and low-mass galaxies, while Bondi (also BH merge) takes over at low redshift and massive galaxies. By dividing the sample into two populations: $\Delta>0$ and $\Delta <0$, we compared their evolution paths following the main progenitors. With evolution tracking, we proposed a simple picture for the BH-galaxy coevolution: Early-formed galaxies seeded black holes earlier, with stellar mass increasing rapidly to quickly reach the point of triggering `jet mode' feedback. This process reduced the cold gas content and stopped the growth of $M_*$, effectively quenching galaxies. Meanwhile, during the initial phase of torque accretion growth, the BH mass is comparable between galaxies formed early and those formed later. However, those galaxies that formed earlier appear to attain a marginally greater BH mass when shifting to Bondi accretion, aligning with the galaxy transition time. As the early-formed galaxies reach this point earlier -- leaving a longer time for them to have Bondi accretion as well as merging, their residuals become positive, i.e., having more massive BHs at $z=0$ compared to these late-formed galaxies at the same $M_*$. This picture is further supported by the strong positive correlation between the residuals and the galaxy age, which we are proposing as a verification with observation data on this story suggested by \simba.}


\maketitle

\section{Introduction}
Galaxies of any substantial size are believed to host a central black hole (BH) with its mass, $M_\bullet$, scaling with the galaxy stellar mass, $M_*$, even in dwarf galaxies \citep[e.g.][]{Mezcua2018a, Reines2020, Birchall2020} and also at extremely high redshift \citep[e.g.][]{Larson2023, Kocevski2023}. The initial states of these BHs, such as the `seeding' mass and mass growth from the very beginning to the massive BHs, are still unknown \citep[e.g.][and many more]{Wise2019, Haemmerle2020, Volonteri2021}. Recent studies revealed that at high redshifts ($z>7$), the BH mass already exceeds $10^8\msun$ \citep[e.g.,][]{Banados2018, Wang2021}, which requires super-Eddington accretion \citep[e.g.][]{Alexander2014}. Along with their mass growth through gas accretion, BHs can release large amounts of energy to galaxies through active galactic nucleus (AGN) feedback, which is also observed in a broad range of redshifts \citep[e.g.,][]{Bongiorno2014, Reines2015, Mezcua2018, Yang2023, Maiolino2024a}. Such a large amount of energy is critical in hydrodynamical simulations and semi-analytic models to regulate star formation \citep[see][and references therein and after]{Heckman2014}, and reproduce the observed galaxy properties \citep[e.g.,][]{Di2005, McCarthy2011, Beckmann2017, Donnari2021, Cui2021}. Whether this energy is responsible for galaxy quenching continues to be a topic of debate in observations; here, we only refer interested readers to many detailed studies, for example, \cite{Aird2012, Page2012, Bongiorno2016, Bluck2020}. 

It has long been realised that the growth of BHs and their host galaxies is correlated, which can be viewed from their properties. Many studies based on observations have found a tight correlation between the BH mass $M_\bullet$ and the stellar mass $M_*$, \citep[e.g.,][]{Magorrian1998, Reines2015, Mountrichas2023}, bulge luminosity $L_{\rm bulge}$ \citep[e.g.,][]{Kormendy1995, Kim2008, Gultekin2009, Sani2011}, bulge mass $M_{\rm bulge}$ \citep[e.g.,][]{McLure2002, Haring2004, Kormendy2013} and stellar velocity dispersion $\sigma$ \citep[e.g.,][]{Merritt2001, Graham2008, McConnell2013, Molina2024}. These scaling relations suggest the coevolution of BHs and their host galaxies, although there is considerable scatter around the median relations. This scatter, more specifically, the residuals in these relations, indicates the intrinsic difference between the BHs at the same galaxy mass. Therefore, to understand the origin of this residual, it is crucial to explore how BHs form and evolve, as well as how they interact with their host galaxies. Through this investigation, we will probe the role of BH in its coevolution with its host galaxy.

In the past few decades, observational research on the $M_\bullet - M_*$ relation has primarily focused on relatively low redshifts \citep[e.g.,][]{Cisternas2011, Sun2015, Ding2020}. However, there are still some clues about BH-galaxy coevolution through scaling relations. For example, \cite{Zhuang2023} measured $\sim 11500$ broad-line AGN at $z \leq 0.35$ and found that galaxies tend to evolve towards the local $M_\bullet - M_*$ relation, i.e., when objects are situated above the relation, they experience a significant increase in $M_*$, resulting in more horizontal evolution. In contrast, when these objects are below the relation, they exhibit a more pronounced growth in $M_\bullet$, evolving vertically. Objects that are aligned with the relation tend to progress along the local relation. Nevertheless, there remains a lack of studies on faint objects and at high redshifts. This situation was dramatically improved after the James Webb Space Telescope (JWST) was launched. Several studies based on the JWST have investigated the $M_\bullet - M_*$ relation at early cosmic times ($z>4$) \citep[e.g.,][]{Kocevski2023, Harikane2023, Maiolino2024b, Stone2024}. Some of them showed that galaxies at high redshift tend to host overmassive BHs compared to the local $M_\bullet - M_*$ relation \citep[e.g.,][]{Ubler2023, Harikane2023, Pacucci2023, Kokorev2023}, although there is an argument that this difference is probably enhanced due to the mass uncertainties and selection effects \citep{Li2025}.

Nevertheless, these scaling relations at a particular redshift are the cumulative results over time, which makes it hard to probe the coevolution between galaxy formation and their hosted central BHs. Meanwhile, the comparisons between different redshifts, which are used to study the coevolution, are also problematic because they are not really tracking the progenitors. Over the last decade, we have been able to understand this numerically through implementing empirical models into cosmological simulations and semi-analytic models \citep[e.g.,][]{Wang2015, Nelson2015, Crain2015, Schaye2015, Dubois2016, Weinberger2017, Bower2017, Pillepich2018, McAlpine2018, Dave2019, Blank2019}. However, from the perspective of simulations, different BH and galaxy formation models make variations in their forecasts of the $M_\bullet-M_*$ relation at $z=0$. From the point of BH mass growth, we expect these processes to affect this relation:
\paragraph{BH seeding mass.} In order to include BH particles in, many simulations \citep[e.g.,][]{Vogelsberger2014, Schaye2015, Dave2019} initialise BH as intermediate mass \citep{Greene2020} in the range of $M_\bullet \sim 10^3-10^5~ \msun$ according to different conditions, such as halo or galaxy masses. Therefore, it is not surprising to see that different BH seeding models noticeably influence the low-mass end of the $M_\bullet-M_*$ relation and would have a substantial impact across all mass ranges at high redshifts \citep[e.g.,][]{Wang2019, Hu2022, Bhowmick2025}.
\paragraph{BH accretion and merger.} In simulations, black hole mass mainly increases through accretion and mergers, which also play an important role in regulating the $M_\bullet-M_*$ relation \citep{Zhu2025}. For the former, most simulations, such as TNG and EAGLE, adopted the commonly used Bondi accretion. While \simba ~additionally includes the torque model, which can bring different BH growth trajectories \citep[e.g.][]{Habouzit2021, Habouzit2022}. Although it is still unclear how BH mergers will contribute to the final mass, with the upcoming gravitational wave observations, we are able to pin down their contributions more accurately. 
\paragraph{AGN feedback.} AGN feedback, which has dramatic energy release to the galactic, halo and even beyond environments, is mostly affecting the galaxy stellar mass in the $M_\bullet-M_*$ relation through regulating galaxy cold gas content and star formation activities \citep[e.g.][]{Ma2022, Liu2025}, especially at the high-mass end \citep[e.g.,][]{Li2020, Habouzit2021}. The AGN feedback, as the key connection between the BH and its host galaxy, plays an important role in their coevolution.

To shed light on the BH-galaxy coevolution, we propose to use the BH mass residuals, i.e., the relative BH mass difference at a given galaxy stellar mass, which can provide a view from a unique point. In this work, we used \simba\ simulation \citep{Dave2019} to explore the origin of the residual in the $M_{\bullet} - M_*$ scaling relation, and its implications for BH-galaxy coevolution. We correlated the residual with the properties of the galaxy and BH properties and investigated how BH accretion and cold gas evolution influence the residual and the growth path of $M_{\bullet} - M_*$.

This paper is organised as follows: In Section \ref{sec:Simba}, we present some basic information about \simba ~simulation and detail its BH accretion and feedback models. In Section \ref{sec:scalingrelation}, we show the residual of the $M_{\bullet} - M_*$ scaling relation and its correlation with the properties of galaxies and BHs. We will study the origin of the residual and BH-galaxy coevolution in Section \ref{sec:co_evolution}, conduct some discussions in Section \ref{sec:discuss} and conclude our main findings in Section \ref{sec:conclusion}.

\section{The \simba ~simulation} \label{sec:Simba}
\simba ~\citep{Dave2019} is a cosmological hydrodynamic simulation based on MUFASA \citep{Dave2016}, running with the meshless-finite-mass mode of the GIZMO code \citep{Hopkins2015}. We focus on the simulation run of m100n1024, with a box size of $100~\hmpc$ on the side, which contains $1024^3$ gas particles and $1024^3$ dark matter particles. The mass resolution is $9.6\times10^7\msun$ and $1.82\times10^7\msun$ for dark matter and gas particles, respectively. The cosmology is based on \cite{Planck2016} with parameters: $\Omega_M = 0.3$, $\Omega_\Lambda = 0.7$, $\Omega_b = 0.048$, $H_0 = 68 \kms \rm Mpc^{-1}$.

In \simba, galaxies are identified using a 6D friend-of-friend (FOF) algorithm with a linking length of 0.0056 times the mean particle separation. The properties of galaxies, such as stellar mass $M_*$, star formation rate (SFR), and metallicity, are characterised by the analysis package CAESAR\footnote{\url{http://caesar.readthedocs.org/}}. Star formation is directly modelled with the relation in \cite{Schmidt1959}, by calculating the molecular gas fraction of the total gas, $f_{\rm H_2}$, following the subgrid prescription of \cite{Krumholz2011}. SFR is calculated as $\rm SFR = \epsilon_* f_{\rm H_2} \rho / t_{\rm dyn}$, where $\epsilon_* = 0.02$ \citep{Kennicutt1998}, $\rho$ is the gas density, and $t_{\rm dyn}$ is the local dynamical time. Star formation occurs when $\rm n_H > 0.13~cm^{-3}$. The \HI ~fraction is calculated based on the prescription in \cite{Rahmati2013}, accounting for the self-shielding effect. The \HI ~and $\rm H_2$ are computed on the fly self-consistently in \simba, with their combined amount representing the total neutral gas.

\simba ~adopts the on-the-fly FOF algorithm to seed BHs in galaxies. BHs will be placed in galaxies when $M_* > \gamma_{\rm BH} \times M_{\rm seed}$, with $M_{\rm seed} = 10^4 \msun/h$ and $\gamma_{\rm BH} = 3 \times 10^5$. After being seeded, BHs will grow through a two-mode accretion model. Torque accretion \citep{Angles2015, Angles2017} is for cold ($T<10^5~K$) gas within the black hole kernel, and the gas inflow rate $\dot M_{\rm Torque}$ is driven by disk gravitational instabilities \citep{Hopkins2011}. Bondi accretion \citep{Bondi1952} is for hot ($T>10^5~K$) non-ISM gas, the gas inflow rate referred to as $\dot M_{\rm Bondi}$. The total accretion rate is calculated as $\dot M_\bullet = (1-\eta) \times (\dot M_{\rm Torque} + \dot M_{\rm Bondi})$, where the radiative efficiency $\eta$ is set to 0.1.

The BH feedback in \simba\ consists of kinetic feedback and X-ray feedback. The kinetic feedback model is motivated by two-mode AGN feedback in observations: the `radiative mode' with high accretion rate, and `jet mode' with low accretion rate, which are divided by the Eddington ratio $f_{\rm edd} \equiv \dot M_\bullet/\dot M_{\rm edd} = 0.2$, with $M_{\rm edd}$ is the Eddington accretion rate: $M_{\rm edd} = \frac{4\pi G M_\bullet m_p}{\eta \sigma_T c}$. `Radiative mode' mimics the radiative AGN winds; the velocity of outflow is based on $M_\bullet$ and typically around $1000~\kms$. While the `jet mode' will arise when $f_{\rm edd}<0.2$ and $M_{\bullet, lim} > 10^{7.5}\msun$. The outflow velocity increases strongly as $f_{\rm edd}$ drops and reaches the full speed $\sim 8000~\kms$ when $f_{\rm edd} \leq 0.02$. When the full-velocity jet is triggered and the gas fraction is $f_{\rm gas} = M_{\rm gas}/M_* < 0.2$, the X-ray feedback is included to additionally heat the surrounding gas. We refer the reader to \cite{Dave2019} for more details.

In this study, we limit the galaxy stellar mass $M_* > 10^{10} \msun$ and BH mass $M_\bullet> 5.0 \times 10^6 \msun$ at all redshifts for analysis to ensure that only well-resolved galaxies and BHs are included.

\begin{figure}
    \centering
    \includegraphics[width=\columnwidth]{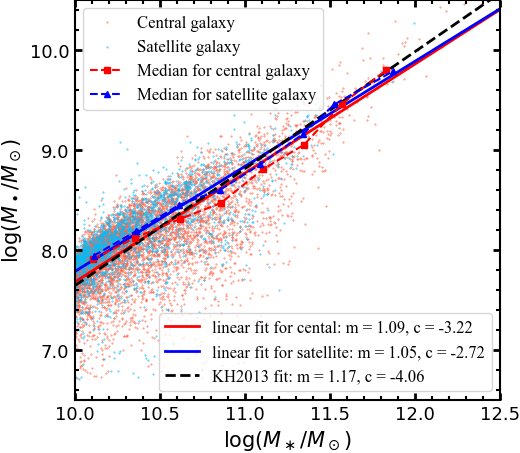}
    \caption{The $M_{\bullet} - M_*$ scaling relation for central and satellite galaxies, which are shown with red and cyan dots, respectively. The best-fit power-law relations for central (red line) and satellite galaxies (blue line) are compared to the observational results from \cite{Kormendy2013} (KH2013, black dashed line), and the best-fit parameters are indicated in the lower-right legend. This plot highlights the very small discrepancy between central and satellite galaxies, as well as the observational relation.}
    \label{fig_cen_sat}
\end{figure}

\begin{figure*}
    \centering
    \includegraphics[width=0.98\textwidth]{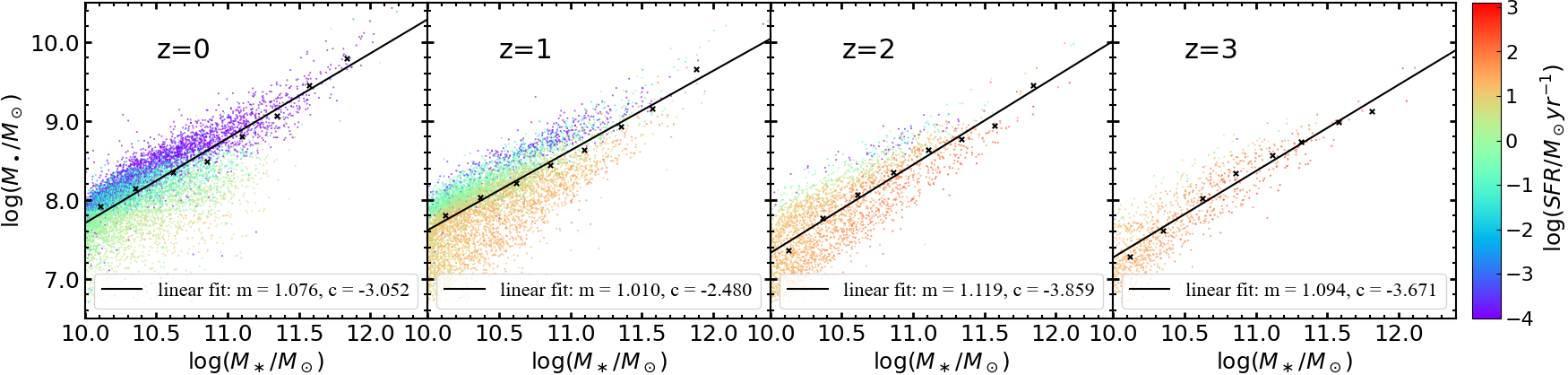}
    \caption{The $M_{\bullet} - M_*$ scaling relation for galaxies at $z=0$, $z=1$, $z=2$ and $z=3$ from left to right panels, colour coded by the $\rm log SFR$. The median $M_{\bullet}$ value in each $M_*$ bin is shown as black dots, with the black lines being the best-fit relation of these median values, which will be used as the median $M_{\bullet} - M_*$ scaling relation for calculating residuals. The best-fit parameters are presented in the bottom legend. Clearly, galaxies with lower SFR at a given $M_*$ tend to host more massive BHs.}
    \label{fig_mbhms}
\end{figure*}

\begin{figure*}
    \centering
    \includegraphics[width=0.94\textwidth]{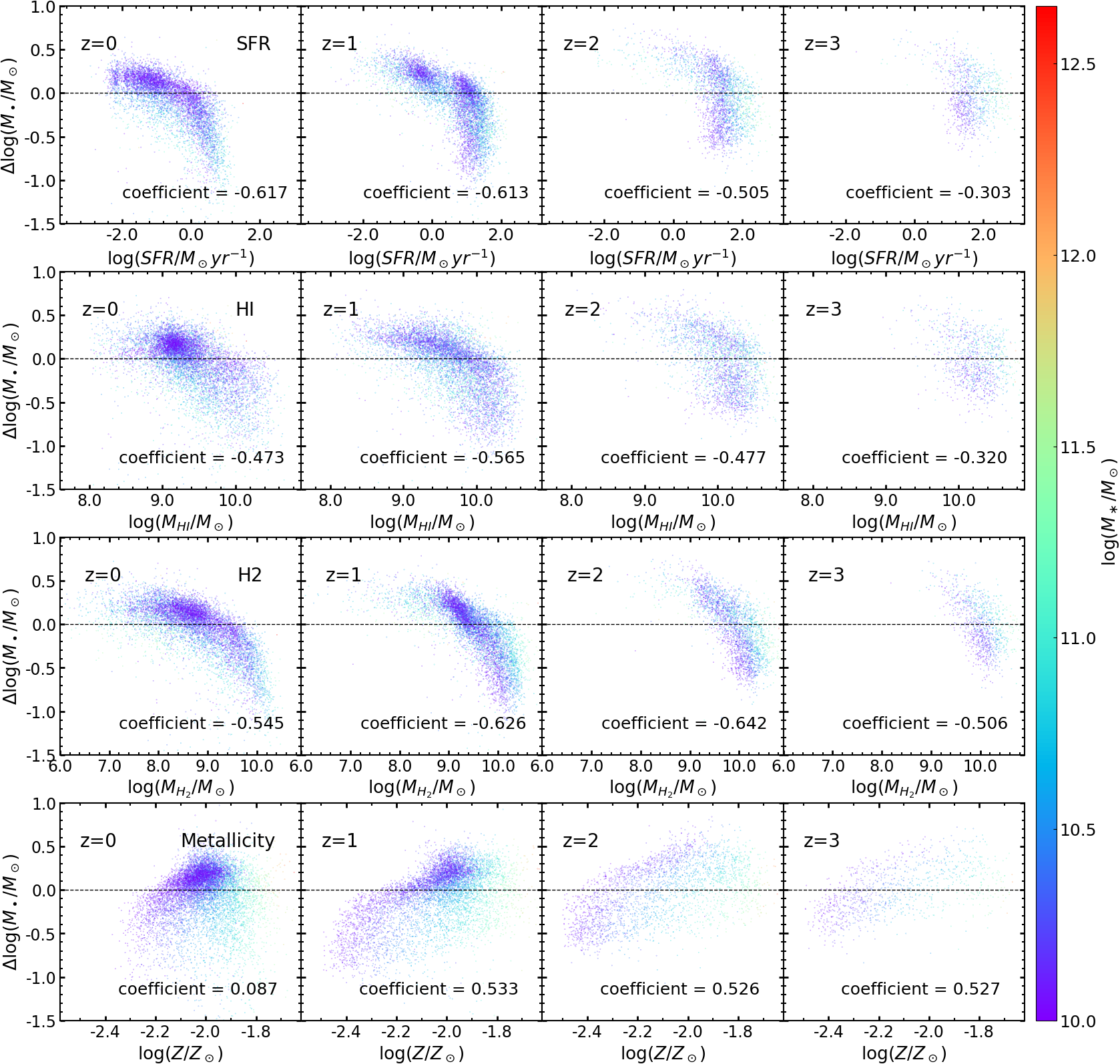}
    \caption{The correlations between residuals and galaxy properties. From top to bottom, we show galaxy SFR, \HI ~mass ($M_\HI$), $\rm H_2$ mass ($M_{\rm H_2}$), and metallicity (Z), respectively, colour coded by their stellar mass ($\log M_*$). From left to right, we show these relations at $z=0$, $z=1$, $z=2$ and $z=3$, respectively. The correlation coefficient values between these relations are shown at the bottom of each panel.}
    \label{fig_sca_galaxy}
\end{figure*}

\begin{figure}
    \centering
    \includegraphics[width=\columnwidth]{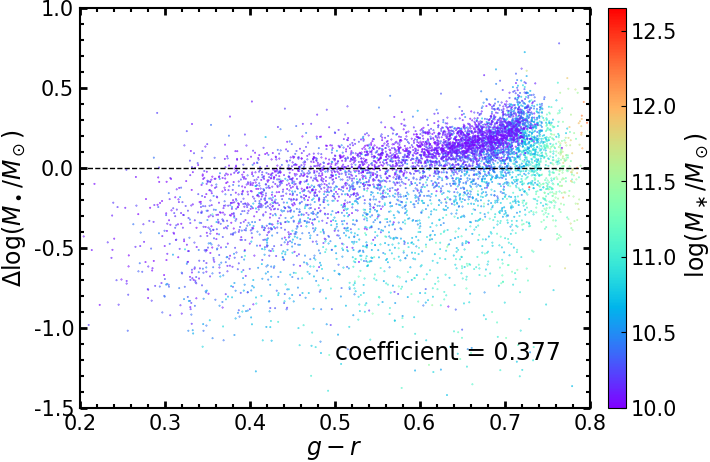}
    \caption{The correlation between residual and galaxy colour $g-r$ at $z=0$, colour coded by $\log M_*$. The correlation coefficient between colour and the residual is shown at the bottom.}
    \label{fig_color}
\end{figure}

\section{The BH -- galaxy scaling relations} \label{sec:scalingrelation}
The essential connection between the mass of a galaxy and its central supermassive BH has been shown to be linearly correlated over orders of magnitude in the logarithmic space in the form of 
\begin{equation}
    \log (M_\bullet/\msun) = m\log(M_*/\msun) + c. \label{eq: mbh_ms}
\end{equation}
Thanks to the torque accretion model \citep{Angles2013, Angles2015, Angles2017}, the \simba ~simulation is able to reproduce this scaling relation, which has been presented in \cite{Dave2019}, as well as other BH and AGN statistics \citep{Thomas2019}. In \autoref{fig_cen_sat}, we revisit this feature by separating central galaxies from satellite galaxies and fitting the $M_\bullet-M_*$ relation using the median values of $M_\bullet$ in each $M_*$ bin. As shown in the figure, the median values and the fitting results for both central and satellite galaxies are very well matched with the observational line from the $M_\bullet-M_{\rm bulge}$ relation in \cite{Kormendy2013} (KH2013). Although the best-fit result for satellite galaxies tends to deviate from centrals as well as observations at low $M_*$ (with slightly higher $M_\bullet$), the median data points are well aligned. One simple explanation could be that these satellite galaxies in massive haloes tend to be stripped to lower $M_*$ compared to the centrals. Because the difference is very small, we will not separate central galaxies from satellite galaxies in this study unless noted.

In \autoref{fig_mbhms}, we show the $M_\bullet - M_*$ relation for galaxies at $z = 0$, $z = 1$, $z = 2$ and $z = 3$ from left to right, colour-coded by $\rm \log SFR$. At a given stellar mass, galaxies tend to have higher BH masses when SFR goes down. Quiescent galaxies host more massive BHs and show a smoother decline in SFR with $M_\bullet$ at $z \le 2$, which is also found in Illustris, IllustisTNG, and EAGLE \citep{Terrazas2016, Terrazas2020}. We fit the median $\log M_\bullet$ in each $\log M_*$ bin (black crosses), the best-fit results are shown as solid black lines, with the linear fitting parameters labelled in the bottom legend. Instead of using the original fit to all the data \citep[e.g.][]{Shankar2025}, we use the median points to avoid the fitting line tilting lower at low galaxy stellar mass because of the large spread of data points in that region.
Our results show that there is a mild evolution from $z = 3$ to $z = 0$. In a given stellar mass bin, more massive BHs are found at lower redshifts. SFRs of massive galaxies above the scaling relation start to decrease earlier than those below the scaling relation, and when $M_\bullet > 10^8~\msun$, galaxies start to quench. This is caused by the `jet mode' AGN feedback in \simba, which we will discuss in the following sections.

As has been shown in \cite{Habouzit2021}, \simba\ agrees with TNG simulations, and tends to have this relation evolve from a lower normalisation to higher values at $z=0$. While Illustris, Horizon-AGN and EAGLE tend to have the opposite redshift evolution, i.e., a higher $M_\bullet-M_*$ normalisation at higher redshift. In the BRAHMA simulation, \cite{Bhowmick2025} explored five different BH seeding models and found that the seeding models would noticeably affect the $M_\bullet-M_*$ relation, especially at the low-mass end and would have a substantial impact across all mass ranges at high redshifts. Although the recent finding of supermassive BHs at very high redshift tends to prefer the decreasing evolution \citep[see][for example]{Ding2020}, as suggested by Illustris, Horizon-AGN EAGLE, and UniverseMachine \citep{Terrazas2024}, larger samples and more consistent studies are needed to reach a firm conclusion on this.

As the colour bar interestingly indicates in \autoref{fig_mbhms}, there is a meaningful scatter around the median $M_\bullet-M_*$ relation. To better quantify this information, we defined the residual $\Delta\log(M_\bullet/\msun)$ ($\Delta$) in the $M_{\bullet} - M_*$ scaling relation as the difference between the actual galaxy $\log (M_{\bullet}/\msun)$ values and their corresponding best-fit $\log (M_{\bullet}/\msun)$ values at the given stellar mass $\log(M_*/\msun)$, i.e.,
\begin{equation}
    \Delta \log(M_\bullet/\msun) = \log(M_\bullet/\msun) - (m \ \log(M_*/\msun) + c).
\end{equation}
We will explore the correlation between the residual and properties of galaxy and BH in detail in Section \ref{subsec:scatter_galaxy} and Section \ref{subsec:scatter_BH}.

\subsection{The relation between residual and galaxy properties} \label{subsec:scatter_galaxy}

\autoref{fig_sca_galaxy} explores the correlations between residuals and various galaxy properties, including galaxy SFR, \HI ~mass ($M_\HI$), $\rm H_2$ mass ($M_{\rm H_2}$) and metallicity (Z), from top to bottom, respectively. From left to right, we show these relations at $z=0$, $z=1$, $z=2$, and $z=3$, respectively. Data points are colour-coded by $\log M_*$. The correlation coefficients between these relations are shown at the bottom of each panel. For all these relationships, we used the Pearson correlation coefficient, which is based on the proportion of covariance to the standard deviation of the initial data. These correlation coefficient values are provided to roughly indicate their relevance and are mostly used for comparisons between different redshifts. Note that we have checked many other galaxy properties; only those with high correlation coefficient values are shown in \autoref{fig_sca_galaxy}. For example, although the SFR is closely related to galaxy colour\footnote{Note that SFR is an instinct values, while colour requires longer time to become distinctive, see the discussions in the supplement of \cite{Cui2021}}, the correlation between galaxy colour and residual is much weaker, as shown in \autoref{fig_color} at $z=0$. The correlations between residuals and galaxy colour tend to become even weaker at higher redshifts.

SFR has a strong negative correlation with the residual, which is consistent with \autoref{fig_mbhms}. In addition, we find that galaxies with higher BH mass (larger $\Delta \log M_\bullet$ ) tend to have smaller cold gas reservoirs (both \HI ~and $\rm H_2$), but higher metallicity and redder colour at $z=0$ \citep[see also][]{Cui2024}. This indicates that star formation activity and cold gas content play a significant role in regulating the $M_{\bullet} - M_*$ relation. Phenomenally speaking, large cold gas reservoirs fuel star formation, maintaining galaxies blue and promoting galaxies to have larger $M_*$ with lower residual, i.e., lying below the relation. When cold gas is consumed/expelled/heated, galaxies start to quench, the growth of $M_*$ slows down, galaxies tend to become redder, and have less cold gas and higher metallicity, which leads to increased and larger residuals, i.e., lying above the relation. Note here that we are simply talking about the effect of galaxy stellar masses on the residual without considering the variations of BH masses, which will be examined in the following section.

\subsection{The relation between residual and BH accretion} \label{subsec:scatter_BH}
Similar to \autoref{fig_sca_galaxy}, we show the correlation between the residual and BH accretion properties at $z=0$, $z=1$, $z=2$, and $z=3$ in \autoref{fig_sca_bh}. From top to bottom panels, we show the BH mass attributed to torque accretion($M_{\rm Torque}$), the torque accretion rate ($\dot M_{\rm Torque}$), the BH mass attributed to Bondi accretion ($M_{\rm Bondi}$), the Bondi accretion rate ($\dot M_{\rm Bondi}$), and the Eddington ratio ($f_{\rm edd}$), respectively, colour coded by $\log M_*$. We show BH accretion rate $\sim 0$ as $10^{-7}$ at each redshift. 

As expected, $M_{\rm Torque}$ and $M_{\rm Bondi}$ are positively correlated with the residual. As we mentioned in Section \ref{sec:Simba}, torque accretion primarily drives black hole growth by accreting cold gas in the vicinity of the black hole, while Bondi accretion facilitates growth through the accretion of hot gas. The increasing masses associated with these accretion modes lead to the growth of black hole mass and a larger residual. We note that residuals of galaxies with similar $M_*$ follow the same linear correlation with $M_{\rm Torque}$, different from the relation of $M_{\rm Bondi}$. This indicates that torque accretion does not significantly impede the growth of $M_*$, suggesting that the growth of the black hole and $M_*$ is synchronised during torque accretion. Therefore, we expect that Bondi accretion plays a more important role in increasing this scatter in \simba, as indicated by the higher correlation coefficients of residuals with $M_{\rm Bondi}$ compared to $M_{\rm Torque}$, especially at lower redshifts. However, the relationship between $M_{\rm Bondi}$ and the residual shows a truncation at $\log (M_{\rm Bondi}/\msun) \sim 5$. We checked that this truncation appears when $\log (M_\bullet/\msun) \sim 8$, at which the `jet mode' feedback is about to start. This indicates that when BHs grow larger, `jet mode' feedback plays a crucial role in heating cold gas. It substantially enhances the Bondi accretion rate and quenches galaxies by increasing the amount of hot gas, thereby modulating the growth of the BH and its host galaxy.

$\dot M_{\rm Torque}$ overall exhibits a negative correlation with residual, while $\dot M_{\rm Bondi}$ shows a positive correlation with residual. However, at $z>0$, for BHs with residuals smaller than 0 ($\Delta<0$), $\dot M_{\rm Torque}$ is positively correlated with residual. In these cases, $\dot M_{\rm Torque}$ is relatively high, and only a small fraction of black holes have initiated Bondi accretion, meaning that black hole growth is primarily driven by torque accretion. In contrast, for those with residual larger than 0 ($\Delta>0$), $\dot M_{\rm Torque}$ is negatively correlated with residual. In these systems, the black hole has grown above the $M_{\bullet} - M_*$ scaling relation, and Bondi accretion increases, leading to rapid black hole growth when torque accretion diminishes. $\dot M_{\rm Bondi}$ is positively correlated with the residual and stellar mass for BHs with $\Delta>0$, suggesting that once Bondi accretion becomes strong, it significantly enhances the mass growth of all BHs.

$f_{\rm edd}$ is defined as $\dot M_\bullet / \dot M_{\rm edd}$, $\dot M_{\rm edd}$ is simply proportional to the BH mass. As the perspective of $\dot M_\bullet$, for over $80\%$ BHs, $f_{\rm edd}$ is dominated by $\dot M_{\rm Torque}$,  which is also the dominant component of $\dot M_\bullet$ seeing from \autoref{fig_sca_bh}. Although $\dot M_{\rm Bondi}$ is positively correlated with the residual, there are fewer than $40\%$ BHs that have $\dot M_{\rm Bondi}>0$ at each redshift, and even $\sim$10\% at $z=3$. These BHs with $\dot M_{\rm Bondi}>0$ are mostly coming from the most massive galaxies at $z=0$.
Therefore, it is not surprising to see that the residuals are anti-correlated with $f_{\rm edd}$, similarly to $\dot M_{\rm Torque}$.  If we naively take $f_{\rm edd}$ as $\dot M_\bullet$, the negative correlation of residuals with $f_{\rm edd}$ and the galaxy SFR will imply a positive correlation between $\dot M_\bullet$ and galaxy SFR, which was also found by previous studies \citep[e.g.,][]{Chen2013, Aird2019}. In \autoref{fig_sca_bh}, we also include $f_{\rm edd} = 0.2$ as the red vertical dashed lines, which denote the starting point of the jet-mode feedback in \simba. It is interesting to see that most ($\sim 95\%$) of BHs entered the jet-mode feedback ($f_{\rm edd}<0.2$ and $M_\bullet > 10^{7.5}~\msun$) after $z\sim 1$, but these systems remain dominated by torque accretion, which naturally produces declining median Eddington ratios at lower redshifts \citep{Angles2015}. The black hole masses with relatively higher values tend to have lower $f_{\rm edd}$. Therefore, on the one hand, this potentially increases the BH mass, especially for low-mass galaxies, as suggested by \cite{Manzano-King2019}; for the AGN outflow is an important and potentially dominant source of feedback in low-mass galaxies, of which the stellar mass growth can be easily ceased to make the $M_{\bullet}$ elevated. On the other hand, it seems not to agree with the AGN activities in dwarf galaxies found in observations. For example, \cite{Mezcua2024} found only $\sim 20$ per cent of dwarf galaxies in MaNGA host AGN activities. 

\subsection{Indication of $M_{\bullet} - M_*$ scaling relation} \label{subsec:indication}
\begin{figure*}
    \centering
    \includegraphics[width=\textwidth]{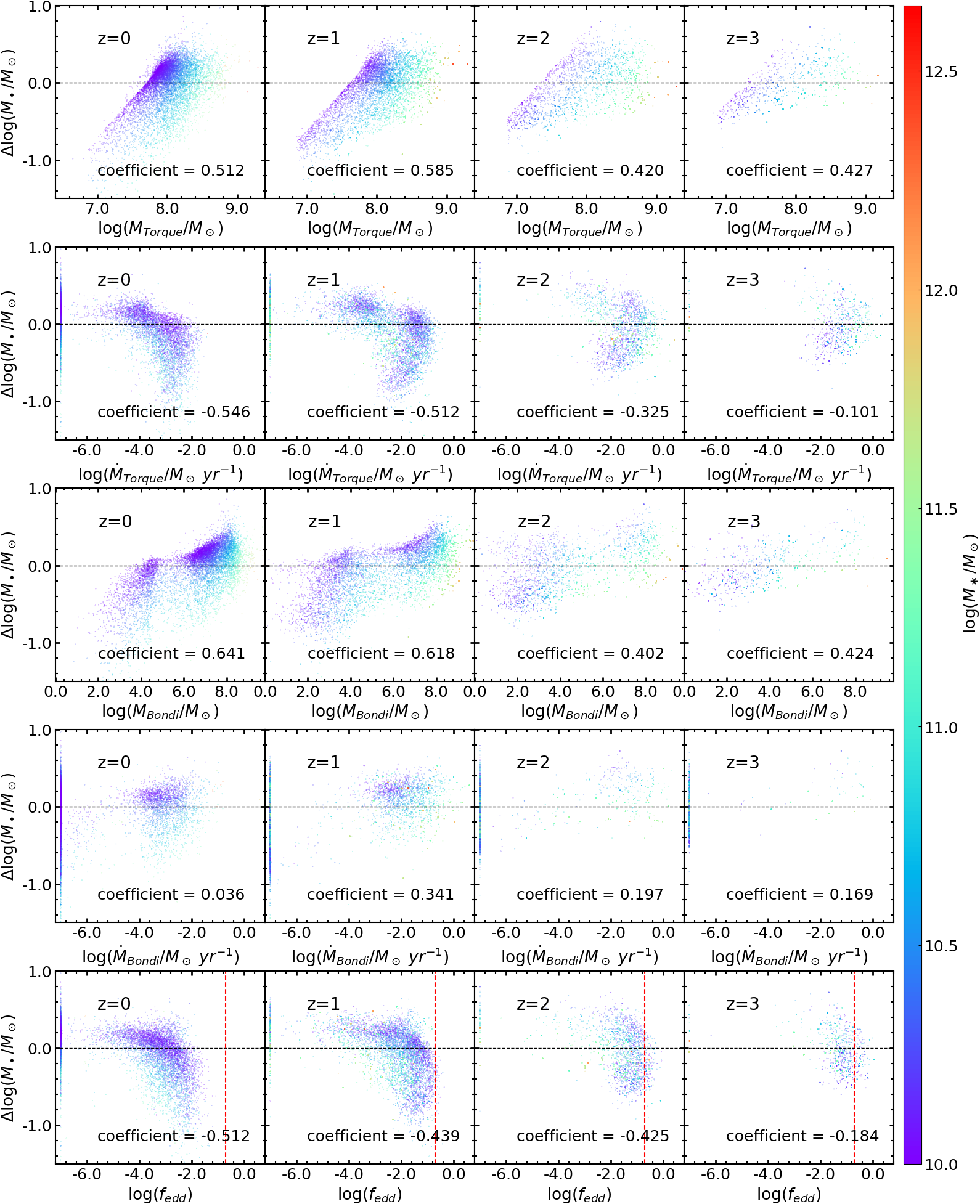}
    \caption{The correlations between the residual and BH properties. From top to bottom, we show BH mass attributed to torque accretion ($M_{\rm Torque}$), the torque accretion rate ($\dot M_{\rm Torque}$), the BH mass attributed to Bondi accretion ($M_{\rm Bondi}$), Bondi accretion rate ($\dot M_{\rm Bondi}$) and Eddington ratio ($f_{\rm edd}$), respectively, colour coded by $\log M_*$. The red dashed lines in the bottom row indicate $f_{\rm edd} = 0.2$, when the `jet mode' feedback is about to start. In the panels for $\dot M_{\rm Torque}$, $\dot M_{\rm Bondi}$, and $f_{\rm edd}$, we represent BHs with zero accretion rate as $10^{-7}$. From left to right, we show these relations at $z=0$, $z=1$, $z=2$, and $z=3$, respectively. The correlation coefficients between these relations are shown at the bottom of each panel.}
    \label{fig_sca_bh}
\end{figure*}

From Figures \ref{fig_mbhms} to \ref{fig_sca_bh}, we have the initial impression that in \simba, the residuals correlate with both galaxy and BH properties. This is a meaningful and interesting quantity for us to interpret and investigate the $M_{\bullet} - M_*$ relation and its evolution. Meanwhile, these properties, such as the cold gas content and BH growth, may be responsible for the residuals in the $M_{\bullet} - M_*$ relation. Through these figures, we can hypothesise that galaxies have a large amount of cold gas at high redshifts, fuelling star formation to grow stellar mass and keep galaxies blue, supplying the torque accretion to grow BH mass quickly as well. Galaxies below the scaling relation should have slightly more cold gas and lower metallicity. When BHs grow to $\sim 10^8~\msun$, the `jet mode' AGN feedback turns on. The bipolar jet ejects gas particles surrounding the central BH vertically out with a high velocity. These gas particles will become wind particles, which first decouple from the hydrodynamic calculation in the simulation to travel almost freely for 0.0001 Hubble time \citep{Dave2019}. Then, it will become a normal gas particle entering the hydrodynamic calculations and heat up their surrounding gas particles through shock heating. This significantly reduces the total cold gas within and around galaxies \citep{Ma2022}. It further promotes BH growth through Bondi accretion. Galaxies then lose cold gas and start to quench, and as a result, the residuals begin to grow. 

This coevolution process is also suggested in observations. \cite{Terrazas2016} found that quiescent galaxies host more massive black holes than star-forming galaxies with similar stellar masses at $z<0.034$, i.e., quiescent galaxies have a positive residual in the $M_{\bullet} - M_*$ relation, in which AGN feedback ceases star formation. Similarly, \cite{Martin-Navarro2018b} also pointed out that BHs above the mean $M_\bullet - \sigma$ scaling relations tend to reside in galaxies with lower star formation rates. With the advantage of simulation, tracking the galaxy and BH evolution over time, we will verify that assumption in the following sections.

\begin{figure*}
    \centering
    \includegraphics[width=\textwidth]{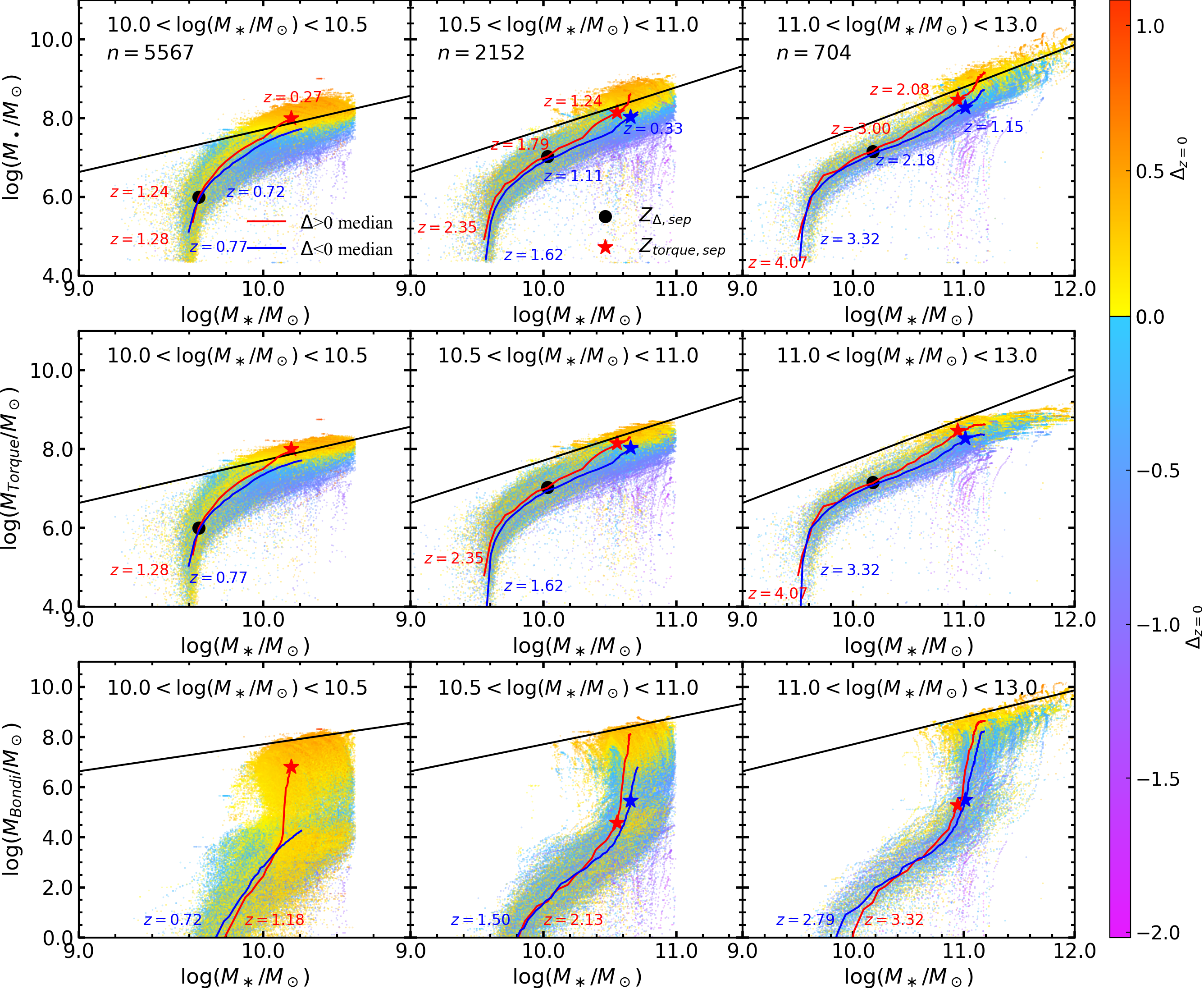}
    \caption{Galaxy and BH coevolution history shown by the tracked $M_*$ and BH masses. From top to bottom, we show the $M_{\bullet}-M_*$, $M_{\rm Torque}-M_*$ and $M_{\rm Bondi}-M_*$ relations as data points colour coded by their residuals at $z=0$ ($\Delta_{z=0}$). Galaxies with stellar mass at $z=0$ in the ranges $10.0<\log(M_*/\msun)<10.5$, $10.5<\log(M_*/\msun)<11$, and $11.0<\log(M_*/\msun)<13$ are shown from the left to right columns, respectively. The galaxy number (n) of each stellar mass bin at $z=0$ is labelled in the top left corner. For reference, we show the scaling relation at $z=0$ from \autoref{fig_mbhms} as the black solid line. We also show the median growth history for galaxies with $\Delta>0$ and $\Delta<0$ as red and blue solid lines, respectively, with the start redshifts of growth labelled. The separation time $Z_{\rm \Delta, \ sep}$ of $M_{\bullet}-M_*$ growth histories of the $\Delta>0$ and $\Delta<0$ populations, defined as the time at which the variance between their median growth trajectories exceeds 0.1 dex, are indicated by large black dots. In addition, we mark the separation time $Z_{\rm Torque,sep}$ between the median $M_{\bullet}-M_*$ and $M_{\rm Torque}-M_*$ with filled stars: red for $\Delta>0$ and blue for $\Delta<0$. To highlight these differences, we also label the redshifts of $Z_{\rm \Delta, \ sep}$ and $Z_{\rm Torque,sep}$ in the first row. All of these labelled redshifts are red for $\Delta>0$ and blue for $\Delta<0$.}
    \label{fig_trace}
\end{figure*}

\begin{figure*}
    \centering
    \includegraphics[width=\textwidth]{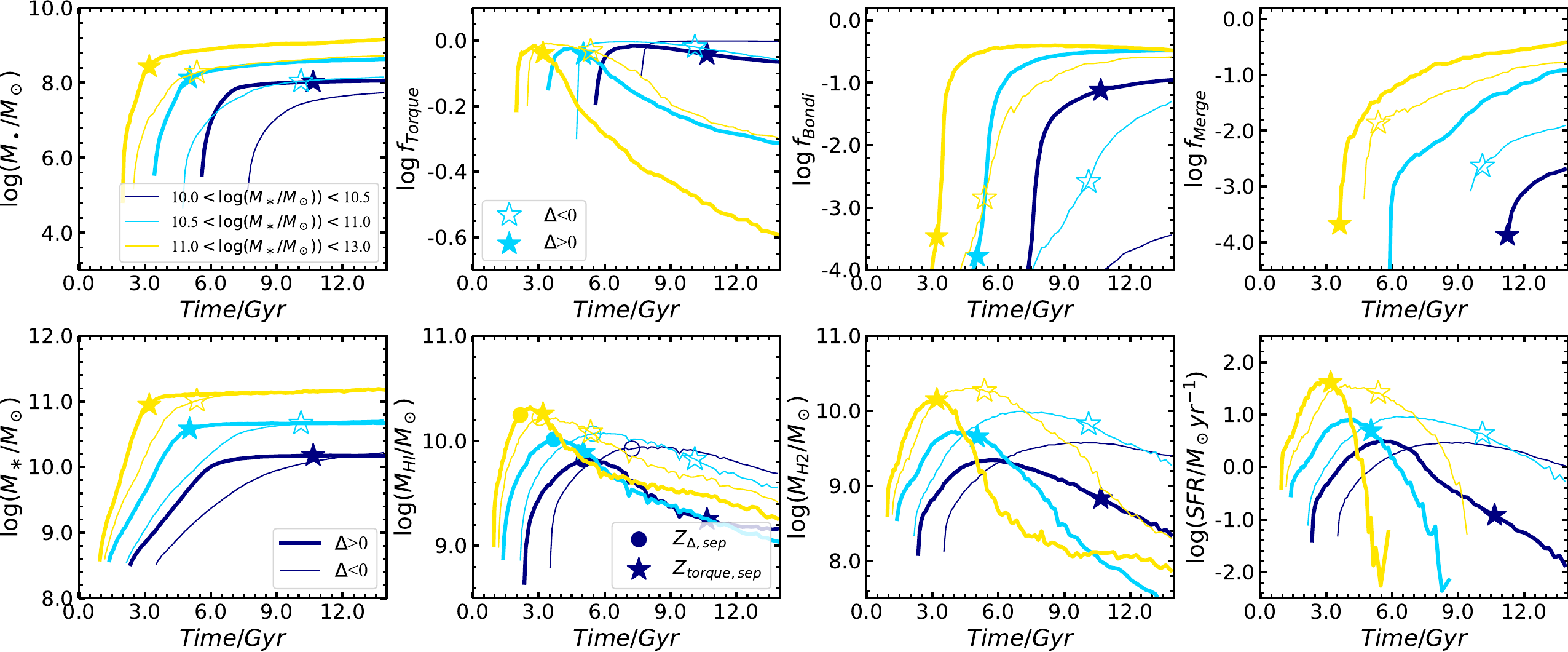}
    \caption{The median growth histories for different galaxy quantities. Top row: the evolution of $M_{\bullet}$, $f_{\rm Torque}$, $f_{\rm Bondi}$ and $f_{\rm Merge}$ with time from left to right, respectively. Bottom row: the evolution of $M_*$, $M_\HI$, $M_{\rm H_2}$ and SFR with time from left to right, respectively. Galaxies with stellar masses in the ranges $10.0<\log(M_*/\msun)<10.5$, $10.5<\log(M_*/\msun)<11$, and $11.0<\log(M_*/\msun)<13$ are depicted using navy, cyan, and yellow lines, respectively. Galaxies with $\Delta>0$ and $\Delta<0$ are represented by thick and thin lines, respectively. We also show the $Z_{\rm \Delta, \ sep}$ (the separation time of $M_\bullet - M_*$ relation for $\Delta>0$ and $\Delta<0$) and $Z_{\rm Torque,sep}$ (the separation time of $M_\bullet - M_*$ and $M_{\rm Torque} - M_*$ relation, see the description in \autoref{fig_trace}) as filled circles and stars for $\Delta>0$ and open symbols for $\Delta<0$, respectively.}
    \label{fig_growth_median}
\end{figure*}

\begin{figure*}
    \centering
    \includegraphics[width=\textwidth]{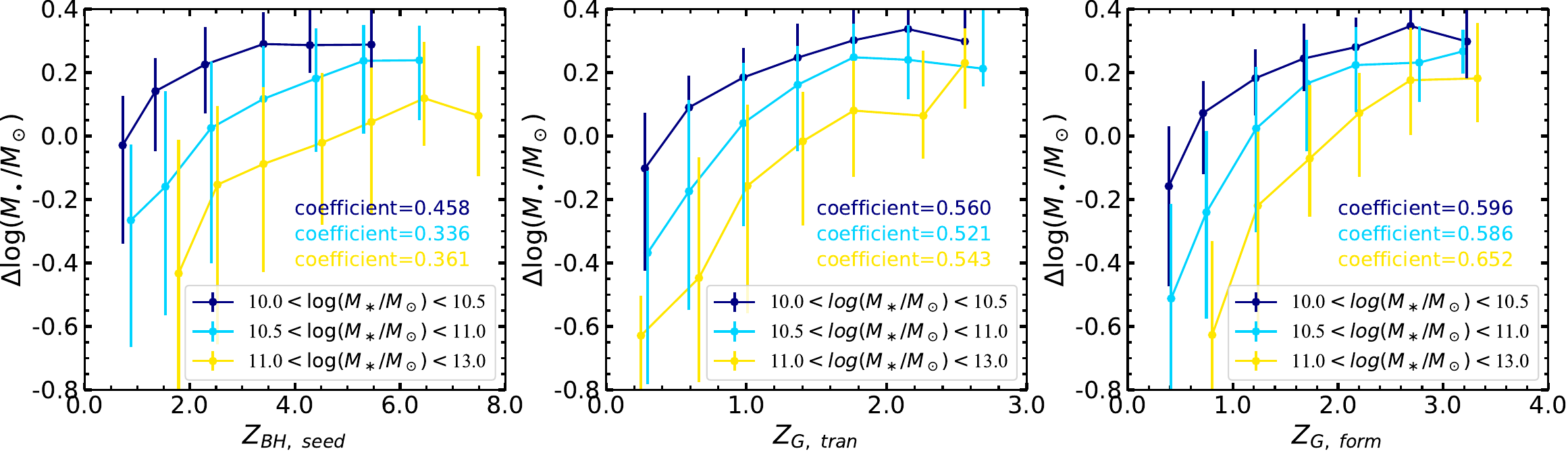}
    \caption{The relation between residuals and the redshifts of BH seeding ($Z_{\rm BH,~seed}$, left panel), galaxy transition ($Z_{\rm G,~tran}$, middle panel), and galaxy formation ($Z_{\rm G,~form}$, right panel). Galaxies with stellar masses in the ranges $10.0<\log(M_*/\msun)<10.5$, $10.5<\log(M_*/\msun)<11$, and $11.0<\log(M_*/\msun)<13$ are depicted using navy, cyan, and yellow lines, respectively, with errors estimated from the 16th to 84th percentile ranges. The correlation coefficients of each stellar mass bin are shown in each panel with their corresponding colours.}
    \label{fig_time}
\end{figure*}

\section{Coevolution of BHs and galaxies}\label{sec:co_evolution}
\subsection{BH and galaxy coevolution} \label{subsec:coevolution}
To further investigate how galaxies and BHs grow together from high redshift to $z=0$ in \simba, we traced the growth of BHs with $\rm M_\bullet > 5.0 \times 10^6~\msun$ and their host galaxies with $\rm M_*> 10^{10}~\msun$ at $z=0$. In \autoref{fig_trace}, we show the distributions of $M_{\bullet}-M_*$, $M_{\rm Torque}-M_*$ and $M_{\rm Bondi}-M_*$ from the top to bottom rows, respectively. The most massive progenitor in the previous snapshot is defined as the main progenitor. Then, each data point in \autoref{fig_trace} marks the position of a progenitor galaxy and its BH in the $M_{\bullet}-M_*$ relation. Thus, the growth path of each galaxy can be traced as a series of dots with the same colour, which is colour coded by the residual at $z=0$ ($\Delta_{\rm z=0}$). Galaxies with stellar masses at $z=0$ in the ranges $10.0<\log(M_*/\msun)<10.5$, $10.5<\log(M_*/\msun)<11$, and $11.0<\log(M_*/\msun)<13$ are shown in the left-to-right columns, respectively. The galaxy number (n) of each stellar mass bin at $z=0$ is labelled in the top left corner. As a reference, we plot the same scaling relation at $z=0$ defined in \autoref{fig_mbhms} as the solid black line. We also show the median growth history for galaxies with $\Delta>0$ and $\Delta<0$ as solid red and blue lines, respectively. To calculate the median lines, we used median stellar and BH masses for each snapshot when there are at least 20 galaxies in each snapshot. In case of a galaxy lacking a BH, the BH mass was considered as 0, and these zero values are included in the calculation of the median.

To more clearly identify when the $M_{\bullet}-M_*$ growth histories of the $\Delta>0$ and $\Delta<0$ populations diverge, we defined a separation time $Z_{\rm \Delta, sep}$, at which the variance between their BH masses of median growth trajectories exceed 0.1 dex for the last time, which is marked with black dots. We also defined the separation time $Z_{\rm Torque, sep}$ when the differences between the median growth trajectories of $M_{\bullet}-M_*$ and $M_{\rm Torque}-M_*$ exceed 0.05 dex, which is marked as filled stars. These stars mark when the total BH mass growth is no longer dominated by the torque mass.

\autoref{fig_trace} does not have redshift information clearly outlined, i.e., the data points at the same position can come from galaxies at different redshifts. To highlight this information, we also include several redshifts in the plot (red for $\Delta>0$ and blue for $\Delta<0$): the redshifts at the bottom indicate the median starting redshift of each median line. Clearly, galaxies with $\Delta>0$ have higher starting redshifts than those with $\Delta<0$ in all three different stellar mass bins. $Z_{\rm \Delta, sep}$ is shown in the middle of the plots in the top row, which agrees with the starting redshift difference. This indicates that the torque model is affected by the galaxy growth history, driving the first deviations in the $M_{\bullet}-M_*$ relation. The $Z_{\rm Torque, sep}$ values are shown on the top of the first row plots, which highlights the influences of the Bondi module on the residuals. Bondi accretion contributes much later but drives the deviations to a larger value. Lastly, this plot also explains why the $M_\bullet - M_*$ relation at high redshift is lower than at low redshift -- the initial growth of BH is below the relation at $z=0$. From the point of the galaxy, the stellar mass growth is much more efficient at high-z compared with low-z, because star formation and the amount of cold gas are systematically decreasing over time due to AGN feedback\footnote{The AGN feedback in \simba\ can affect the gas properties far outside of the halo, we refer to \cite{Yang2024} for the interesting findings.}. As such, low-z galaxies are more affected by the growth of $M_*$, and the $M_{\bullet}-M_*$ relation is higher at low redshift. 

In all stellar mass bins, the growth of the $M_{\bullet}-M_*$ relation for galaxies with either $\Delta>0$ or $\Delta<0$ almost overlapped before $Z_{\rm \Delta, sep}$. In fact, almost all data points lie in the same region without clear separation until a very late redshift. The subtle difference can only be seen through the median lines. Then galaxies with $\Delta>0$ have higher BH masses than those with $\Delta<0$. Before $Z_{\rm Torque, sep}$, the $M_{\bullet}-M_*$ relation closely follows the $M_{\rm Torque}-M_*$ growth trajectory for all galaxies. After $Z_{\rm Torque, sep}$, i.e., when $z \sim 0$ and $\log(M_\bullet/\msun)\sim 8$, `jet mode' AGN feedback is triggered. It expels cold gas from galaxies and heats the cold gas within them. This process promotes Bondi accretion rapidly, which is implemented for hot gas. The significant growth of $M_{\rm Bondi}$ causes the fast growth of $M_{\bullet}$. At the same time, torque accretion slows down, causing the separation of $M_{\bullet}-M_*$ and $M_{\rm Torque}-M_*$. As a consequence, the $M_\bullet-M_*$ growth trajectory vertically moves toward the scaling relation at $z=0$ for galaxies with $\log(M_*/\msun)>10.5$. Galaxies that start Bondi accretion earlier move above the scaling relation at $z=0$ and become quenched. For galaxies with $\log(M_*/\msun)>11$ and $\Delta>0$, the growth of $M_{\rm Bondi}$ slows down when the redshift approaches 0. The growth of these massive galaxies is also influenced by mergers, which further contribute to the increase of BH and stellar masses \citep[e.g.,][]{Hirschmann2015, Habouzit2021}.

\subsection{Physical driver of the residual} 
\autoref{fig_trace} reveals the general $M_{\bullet}-M_*$ evolution trend -- BHs and galaxies simultaneously increase towards the relation at $z=0$, especially at high redshift. That plot highlights the information on the initial state, which depends on the simulation resolution and the BH seeding mass. At lower redshifts, it is not clear when BH and galaxy stellar masses become stable, i.e., no clear evolution results in the intermediate phase of growth trajectory in \autoref{fig_trace}. Furthermore, the distinct mass assembly phases driven by accretion mechanisms and feedback processes are inferred only by several hypotheses. Therefore, what factors lead to the different mass assembly phases, and how can we quantitatively characterise these individual evolutions? More importantly, is the same mass growth of BH and galaxy driving the similarity in the $M_\bullet-M_*$ relations between different redshifts? In the following, we want to investigate deeper into how BHs accumulate mass via accretion processes and mergers, along with examining galaxy growth through star formation and the evolution of their cold gas reservoir as a function of time. 

In \autoref{fig_growth_median}, we plot the growth history of BHs and galaxies. In the top row, we show the median evolution of $M_{\bullet}$, and the mass fractions of torque accretion ($f_{\rm Torque} = M_{\rm Torque}/ M_\bullet$), Bondi accretion ($f_{\rm Bondi} = M_{\rm Bondi}/ M_\bullet$) and BH mergers ($f_{\rm Merge}= M_{\rm Merge}/ M_\bullet$) over time from left to right, respectively. The evolution of $M_*$, $M_\HI$, $M_{\rm H_2}$ and SFR is shown in the bottom row. Galaxies with different stellar mass bins are depicted using different colours. Galaxies with $\Delta>0$ and $\Delta<0$ are represented by thick and thin lines. We also marked $Z_{\rm \Delta, \ sep}$ and $Z_{\rm Torque,sep}$ in the panels.

\autoref{fig_growth_median} shows that BHs with $\Delta>0$ start their growth earlier than those with $\Delta<0$. This is also true for their host galaxies, which is simply due to the seeding scheme in \simba -- BHs are seeded based on the host galaxy mass. As such, early-formed galaxies will have BH seeded earlier. For BHs in galaxies with $\log(M_*/\msun)<11$, torque accretion ($f_{\rm Torque}$) dominates the growth of BH mass, contributing a larger fraction than Bondi accretion ($f_{\rm Bondi}$) or mergers ($f_{\rm Merge}$). Specifically, BHs in galaxies with $10<\log(M_*/\msun)<10.5$ and $\Delta<0$ rely almost entirely on torque accretion. As time passes, the relative contribution of torque accretion diminishes, while Bondi accretion and mergers become increasingly significant. For the most massive systems ($11<\log(M_*/\msun)<13$), Bondi accretion and mergers surpass the contribution from torque accretion and dominate the BH growth at $z=0$. It seems that the merger contribution tends to become more significant slightly later than the Bondi accretion for BHs with $\Delta>0$.  

On the galaxy side, which is shown in the bottom panels of \autoref{fig_growth_median}, the stellar mass growth is primarily driven by star formation from cold gas reservoirs ($M_{\rm \HI}$ and $M_{\rm H_2}$). In the early Universe, the rapid $M_*$ growth is due to the large amount of cold gas and high SFRs. In \autoref{fig_trace}, we can see that galaxies with $\Delta>0$ and $\Delta<0$ share similar $M_{\bullet}-M_*$ growth trajectories before $Z_{\rm \Delta, \ sep}$, which corresponds to the time when $M_{\HI}$ reaches the peak shown in the second panel of the bottom row in \autoref{fig_growth_median}. Subsequently, for galaxies in the range of $\log(M_*/\msun)<11$, those with $\Delta>0$ experience a faster decline in $M_{\HI}$ after reaching the peak, slowing down the torque accretion and the growth of $M_*$ earlier compared to those with $\Delta<0$. However, for the most massive galaxies ($11<\log(M_*/\msun)<13$), those differences between $\Delta>0$ and $\Delta<0$ become smaller, always have similar $M_{\HI}$ and experience the substantial merger, their $M_{\bullet}-M_*$ growth history does not show big differences, as we also showed in \autoref{fig_trace}. \HI ~locates at the outskirts of galaxies, acting as the supply of $\rm H_2$ and revealing the total cold gas reservoir of galaxies \citep[e.g.,][]{Wong2002, Kennicutt2007, Bigiel2008}. As \simba\ uses an $\rm H_2$-based star formation model, it is not surprising to see that the SFR evolution tightly (bottom right panel) follows the $\rm H_2$ mass evolution. The longer and higher SFR in these $\Delta < 0$ galaxies results in higher stellar mass at $z=0$, and their BH mass is still smaller compared to these galaxies with $\Delta>0$ at the same $M_*$. Lastly, we checked that galaxies with $\Delta > 0$ reside in a denser environment than those with $\Delta < 0$ after $Z_{\rm \Delta, \ sep}$, which would deplete their \HI ~more quickly. For the underlying physical mechanism behind this fast \HI\ depletion, we suspect that it is due to (1) the environmental effect, such as ram pressure stripping or tidal events \citep[e.g.,][]{Brown2017, Cortese2021, Wang2021a, Wang2022}, or (2) the earlier, as such longer, AGN feedback, which will be detailed in the coming paragraph, to heat the gas to reduce the \HI\ mass.

Before $Z_{\rm Torque, sep}$, $\rm H_2$ and SFR increase continuously, keeping torque accretion dominated and the consistent growth of $M_{\bullet}-M_*$ and $M_{\rm Torque}-M_*$ shown in \autoref{fig_trace}. When cold gas reservoirs and SFRs reach the peak, that is, after $Z_{\rm Torque,sep}$, the BH mass has already reached $10^8\msun$, as we discussed before, `jet mode' feedback is triggered, ejecting cold gas and increasing the density of hot gas within subhaloes. The reduced cold gas suppresses the torque accretion and star formation, gradually quenching galaxies. Meanwhile, BH growth starts to be increased by Bondi accretion. Then, the growth trajectories of $M_{\bullet}-M_*$ and $M_{\rm Torque}-M_*$ start to separate, which is marked by the stars. In galaxies with $\Delta>0$, BHs experience more significant and earlier Bondi accretion, along with more mergers, leading to overmassive BHs. Note that the BH mergers have similar contributions as Bondi accretion at $z=0$ for the most massive galaxies. Both contributions decrease with galaxy stellar mass, with the BH mass in low-mass galaxies still dominated by torque accretion. Meanwhile, galaxies that host undermassive BHs ($\Delta<0$) experience a longer duration and slowly decrease in star formation, stopping the growth of $M_*$ at a later stage compared to galaxies with overmassive BHs, despite having similar SFRs in the early Universe, which is also found in \cite{Martin-Navarro2018a} and suggested in \cite{Cui2021}. 

We note that most galaxies with $\Delta>0$ start to grow earlier than those with $\Delta<0$, as indicated by the redshifts displayed at the bottom of each panel in \autoref{fig_trace}. This provides a longer time for galaxies with $\Delta>0$ to form stars and accumulate the torque and Bondi accretion to grow BHs more significantly than the $\Delta<0$ population, and therefore causes the larger residual. The different residuals of galaxies are partially attributed to differences in the times of galaxy formation. To quantify the relation between growth time and residual, we fit the galaxy formation history as an error function and define the transition time ($Z_{\rm G, tran}$) for each galaxy following the Equation. 1 of \cite{Cui2021}, which is the redshift when their growth slope is less than 0.1, i.e.,
\begin{equation}
    \frac{\rm d\log M_*}{\rm dt(Gyr)} < 0.1 .
    \label{eq:tran}
\end{equation}
This $Z_{\rm G, tran}$, not surprisingly, aligns with the $Z_{\rm Torque, sep}$, which can be viewed in \autoref{fig_growth_median}. Lastly, we also included one observable, the mass-weighted galaxy age, which is also tightly connected with the galaxy transition time. To have a consistent comparison, the galaxy mass-weighted age is converted to the redshift through the Universe's look-back time. With this quantity, which can be measured observationally, the \simba\ proposed picture of $M_\bullet-M_*$ coevolution can be verified with observational results.

In \autoref{fig_time}, we show the correlations between the residuals $\Delta$ and the redshift when BH seeded ($Z_{\rm BH, seed}$), galaxy transition time ($Z_{\rm G,tran}$) and galaxy age (for uniformity, we convert the galaxy age to the redshift when they formed, $Z_{\rm G, form}$) from left to right panels. Again, the same three galaxy stellar mass bins: $10.0<\log(M_*/\msun)<10.5$, $10.5<\log(M_*/\msun)<11$, and $11.0<\log(M_*/\msun)<13$, are presented in three different colour lines as indicated in the legends. To highlight the significance, we further include the correlation coefficient between $\Delta$ and these quantities for each mass bin in the middle of each plot. The positive correlations between $\Delta$ and the three quantities confirm our previous claims about why some galaxies host low-mass BHs while others can nurture massive BHs.

However, the correlation between residuals and $Z_{\rm BH, seed}$ seems to be saturated when $Z_{\rm BH, seed}$ is high and $\Delta > 0.2$. The correlation curve plateaus at slightly different redshifts for different mass bins -- a little higher at larger mass bins. This implies in two aspects: (1) that there may be a maximum BH mass of the galaxy at a given mass, which cannot be exceeded, even if it is seeded at the highest possible redshift. As the BH mass residual is anti-correlated with the galaxy SFR, these galaxies with high $Z_{\rm BH, seed}$ are formed considerably earlier and get quenched earlier. After the quench, both galaxy stellar mass and BH mass almost halt as shown in the left panels of \autoref{fig_growth_median}. (2) The plateau also indicates that above a certain $Z_{\rm BH, seed}$, even the early-formed galaxy will not push the BH mass even higher, this implies that the coevolution between BH and galaxy after quench becomes self-similar: either both BH and galaxy masses increase fully shut down, or they both increase along the $M_\bullet-M_*$ relation.

The effects of BH seeding mass, especially on the scatter of the $M_\bullet-M_*$, have also been presented in many studies \citep[e.g.,][]{Habouzit2021, Zhu2025}. As discussed in \cite{Habouzit2021}, the BH seeding mass should affect these simulations with only Bondi accretion model ($\dot M_\bullet \propto M^2_\bullet$ more than \simba, which has the torque accretion only weakly depending on the BH mass as the form of $\dot M_\bullet \propto M^{1/6}_\bullet$ \citep{Angles2013, Angles2015}). This does not contradict what we are showing -- the final BH mass is related to the BH seeding time, not the seeding mass. Furthermore, we show that the scatter is contributed by both accretion models, with the Bondi model, as well as mergers, contributing more for massive galaxies and low redshift \citep[see e.g.][]{Zou2024}. 

Since BH seeding in \simba\ is based on the galaxy stellar mass, early BH seeding time means an early-formed galaxy, as such a higher galaxy transition redshift. Following \cite{Cui2021}, at the same stellar mass, galaxies with higher transition redshift tend to live in a late-formed halo, i.e., a smaller halo at high redshift. This makes the cold gas easily be consumed, transferring the BH accretion to Bondi mode. The higher BH mass at the galaxy transition time and longer Bondi accretion period further lead to a higher BH mass to result in the positive residuals. The relation between Bondi accretion and BH mergers can be viewed as mutually reinforcing, for example, \cite{Bhowmick2020} suggested that merging BHs tend to have high Eddington ratios, i.e., higher BH accretion rate. 

All these assumptions are supported by the previous figures. However, to confirm this in observations, we need to transfer this information to some observables. Given that early transition time leads to red galaxies \citep{Cui2021}, it is not surprising to see the positive correlation between $\Delta$ and $Z_{\rm G, form}$. With that, our theoretical prediction can be verified with observational data. As suggested by \cite{Habouzit2022}, the BH mass offsets of the simulated faint quasar population at $z \geq 4$ represent the BH mass offsets of the entire BH population, for all simulations, which can be used to present the entire BH offsets.

\section{Discussion} \label{sec:discuss}

\subsection{The scatter and residuals}
In observation, the scatter in the $M_* - M_\bullet$ relation is often related with the observation uncertainties, systematics and quantity deriving methods, etc. Observations generally giving the scatter larger than 0.5 dex, \citep[see][for example]{Graham2012,Davis2018,Baron2019,Li2023}, the scatter from simulation is normally smaller than 1 dex \citep{Habouzit2021, Zhu2025}. On the one hand, simulations are probing the intrinsic scatter, which is expected to be smaller than observation; on the other hand, that could be related to the lack of stochasticity in the galaxy/BH subgrid. In agreement with \cite{Zhu2025}, they found that the scatter in low-mass galaxies can be significantly reduced by AGN feedback, while at the high-mass end, this is decreased by the hierarchical merging of quenched systems. This was also found by \cite{Hirschmann2010}, who showed that the intrinsic scatter of the $M_\bullet - M_{\rm bulge}$ relation decreases with an increasing number of mergers and so with time. However, these scatter studies can only reflect some systematics, which makes it hard to provide useful information on the BH-galaxy relation. Unlike scatter, which only quantifies the standard deviations, the residual can be directly linked to the different BH mass growth, allowing us to ask the question: why can some BHs accrete their masses more than others at the same $M_*$?

It is well-known in observation that the $M_\bullet-M_*$ relation has different slopes when different galaxy types are used to do the fitting function \citep[e.g.][for more recent results]{Davis2019, Graham2023}. \cite{Shankar2025} found that the correlation between the residuals of the BH scaling relations ($M_\bullet-M_{\rm bulge}$, $M_\bullet-\sigma$) may be the result of kinetic AGN feedback. However, increasing the kinetic output of AGN feedback does not improve the alignment with observational data. \cite{Marasco2021} investigated the residuals in the galaxy global star formation efficiency $f_* - M_\bullet$ relation versus those in the $M_\bullet - M_{\rm halo}$ relation and found that galaxies living in lighter haloes have larger $f_*$ at the same $M_{\bullet}$. This is not directly presented in our study, but is in qualitative agreement if we combine the findings from \cite{Cui2021}, which indicated that blue galaxies generally inhabit lower-mass haloes, with the results shown in \autoref{fig_mbhms}, which demonstrate that massive galaxies are more likely to be star-forming at a fixed BH mass. In addition, by combining the results from \cite{Cui2021}, we expect that low-mass BHs tend to live in low-mass haloes at the same galaxy stellar mass\footnote{Note that we further directly verified this with the simulation data.}. That theoretically explains why the $M_\bullet - M_{\rm halo}$ relation is also very tight. Nevertheless, directly linking this residual to galaxy properties provides a novel insight into BH growth and how it connects with galaxy evolution. Furthermore, the BH mass residuals computed for the full BH population are crucial to understand the build-up of BHs \citep{Habouzit2022}, which we will discuss in the next subsection. 

\subsection{The coevolution viewing from the residual picture}
In \cite{Li2024}, the BH growth from TNG simulation is separated into 4 different phases. As shown in their Figure 2, there are few differences in the cumulative distributions between phase 1 (star formation, BH seed dominates) and phase 2 (rapid SMBH growth), as well as between phase 3 (self-regulation and galaxy quenching) and phase 4 (mergers). A similar feature can also be found in \simba, but for simplicity, we merge the two initial phases as the torque growth stage and the last phases as Bondi and merger stage, which is adopted similarly in \cite{Mo2024}'s model \citep[see also][]{Boco2023}. The first stage is, of course, dominated by torque accretion. At the beginning of galaxy formation, the amount of cold gas rapidly increases, which can supply star formation and rapid BH growth through torque accretion. During this time, galaxies formed earlier have more cold gas in the early Universe, growing $M_*$ and $M_{\bullet}$ faster, making them have higher stellar and BH masses than those galaxies formed later after $Z_{\rm \Delta, \ sep}$.  The second stage starts at $Z_{\rm Torque, sep}$. When the BH mass increases to $\sim 10^8 \msun$, `jet mode' feedback starts, which rapidly decreases and heats the cold gas. The BH growth supplied by torque accretion slows and starts to be dominated by Bondi accretion. The growth trajectories of $M_{\bullet}-M_*$ and $M_{\rm Torque}-M_*$ separate. \HI ~is consumed, $\rm H_2$ and SFR begin to decrease. Galaxies experience a deceleration in $M_*$ growth and initiate quenching. Black holes seeded later result in a slower decline of cold gas and SFR, allowing for a longer period of star formation and a smaller fraction of Bondi accretion, maintaining $\Delta<0$. Furthermore, the change in the neutral hydrogen gas mass (including both $\rm H_2$ and \HI) with the star formation rate of massive central disk galaxies is suggested as a critical constraint of black hole feedback models across several simulations \citep{Shi2022}.

\section{Conclusions} \label{sec:conclusion}

In this work, we used the \simba\ simulation \citep{Dave2019} to extend the studies of \cite{Thomas2019} by detailing our understanding of the residuals in the $M_\bullet - M_*$ relation. Based on the successful BH accretion and AGN feedback models, the \simba\ simulation is able to reproduce the observed $M_\bullet - M_*$ relations. We focus on galaxies that are well resolved in the \simba ~simulation with $M_*>10^{10}~\msun$ and $M_\bullet > 5.0\times 10^{6}~\msun$. Our main findings are as follows.
\begin{itemize}
    \item[1.] \simba ~simulation is able to reproduce the observed $M_\bullet - M_*$ relations of \cite{Kormendy2013}. There is little difference between the satellite and central galaxies for the $M_\bullet - M_*$ relation. The evolution of $M_\bullet - M_*$ relation is weak from $z=0$ to $z=3$ in \simba, with a slight increase in the normalisation.
    
    \item[2.] Residuals, denoted as the relative difference between the galaxy $M_\bullet$ and the $M_\bullet-M_*$ relation, are correlated with several galaxy and BH properties. More specifically, residuals are anti-correlated with the galaxy SFR, $\rm H_2$, \HI ~masses, BH $\dot M_{\rm Torque}$ and $f_{\rm edd}$, and correlated with the galaxy metallicity, colour, BH $M_{\rm Torque}$, $M_{\rm Bondi}$ and $\dot M_{\rm Bondi}$. These correlations indicate that the residuals are a meaningful indicator of the BH-galaxy coevolution instead of random noise.
    
    \item[3.] We investigate the drivers of residuals by tracking the BH and galaxy growth. We defined the separation time $Z_{\rm \Delta,~sep}$ when the $M_\bullet - M_*$ growth trajectories of galaxies with $\Delta>0$ and $\Delta<0$ separate and $Z_{\rm Torque, sep}$ when the BH mass deviates from the torque mass. 
    We find that $Z_{\rm \Delta,sep}$ correlates with the time when $M_{\HI}$ reaches the peak. After $Z_{\rm \Delta,sep}$, galaxies with $\Delta>0$ experience a faster decline in $M_{\HI}$ than those with $\Delta<0$, resulting in a decrease of the total cold gas content and the $M_*$ growth. The $Z_{\rm Torque, sep}$ is related to the time when SFR and $M_{\rm H_2}$ reach peaks. At this time, $M_\bullet$ is up to $10^8~\msun$, `jet mode' AGN feedback turns on, and the amount of cold gas decreases, gradually suppressing the torque accretion, quenching galaxies, and promoting the Bondi accretion, leading to the separation of $M_\bullet - M_*$ and $M_{\rm Torque} - M_*$. After that, BHs growth is dominated by Bondi accretion and merger. With galaxies stopping their stellar mass growth after quenching, the relation begins to grow vertically.
    
    \item[4.] Our results reveal that the BH-galaxy growth history in \simba ~ consists of two stages. The first stage is dominated by torque accretion. The amount of cold gas rapidly increases, which supplies star formation and the rapid BH growth through torque accretion. The second stage starts at $Z_{\rm Torque, sep}$. The BH growth supplied by torque accretion slows and starts to be dominated by Bondi accretion, which is caused by the `jet mode' AGN feedback.
    
    \item[5.] The residuals are also partially attributed to differences in $Z_{\rm G, tran}$, $Z_{\rm G, form}$ and $Z_{\rm BH, seed}$. Galaxies that form earlier tend to have massive BHs and also be quenched earlier. The galaxies that formed stars earlier accumulated greater torque mass, owing to the abundance of cold gas, and gained more Bondi mass (including merged mass) because of a longer period of accretion. Consequently, their BH masses are higher than those of galaxies that formed later, leading to a positive residual.

\end{itemize}

Our results are limited to the \simba\ simulation, which would be interesting to verify with other simulations with different baryon models. More interestingly, do observational results agree with \simba\ predicted correlations between the BH residuals and their host galaxy colours/ages? Although the BH mass estimated in observation is very uncertain and relies on different assumptions, the residuals, the relative difference, should be free from these systematics and assumptions. Therefore, this measurement is more robust and can be used for probing the BH-galaxy coevolution if proved to be true. Lastly, we do not see any clear differences at the galaxy cluster scale, $M_* \gtrsim 10^{12}$, due to the number statistics. We are currently working with The Three Hundred galaxy clusters \citep{Cui2018b, Cui2022} to investigate this $M_\bullet-M_*$ relation in more detail by studying the residual correlations with galaxy cluster properties.

\begin{acknowledgements}
We thank the reviewer for the thorough and thoughtful comments of our paper. The authors would like to thank James Aird, Antonis Georgakakis, Beatriz Mingo, Andrea Merloni, Johannes Buchner and others in the AGN survey workshop for useful discussions, and especially thank the organisers for such a wonderful workshop. We gratefully thank Philip Hopkins for making Gizmo public, and providing our group with early access.

This work is supported by the National SKA Program of China (grant No. 2020SKA0110100), the CAS Project for Young Scientists in Basic Research (No. YSBR-092), China Manned Space Program (grant No. CMS-CSST-2025-A04), GHfund C(202407031909) and the Munich Institute for Astro-, Particle and BioPhysics (MIAPbP), which is funded by the Deutsche Forschungsgemeinschaft (DFG, German Research Foundation) under Germany´s Excellence Strategy – EXC-2094 – 390783311. We acknowledge the use of the High Performance Computing Resource in the Core Facility for Advanced Research Computing at the Shanghai Astronomical Observatory. WC gratefully thanks Comunidad de Madrid for the Atracci\'{o}n de Talento fellowship No. 2020-T1/TIC19882 and Agencia Estatal de Investigación (AEI) for the Consolidación Investigadora Grant CNS2024-154838. He further acknowledges the Ministerio de Ciencia e Innovación (Spain) for financial support under Project grant PID2021-122603NB-C21, ERC: HORIZON-TMA-MSCA-SE for supporting the LACEGAL-III (Latin American Chinese European Galaxy Formation Network) project with grant number 101086388. D.A.A. acknowledges support from NSF CAREER award AST-2442788, an Alfred P. Sloan Research Fellowship, and Cottrell Scholar Award CS-CSA-2023-028 by the Research Corporation for Science Advancement. 

The \simba ~simulation was run on the DiRAC@Durham facility managed by the Institute for Computational Cosmology on behalf of the STFC DiRAC HPC Facility. The equipment was funded by BEIS capital funding via STFC capital grants ST/P002293/1, ST/R002371/1, and ST/S002502/1, Durham University, and STFC operations grant ST/R000832/1. DiRAC is part of the National e-Infrastructure.

This work has made extensive use of the Python packages --- Ipython with its Jupyter notebook \citep{ipython}, LMFIT \citep{LMFIT}, NumPy \citep{NumPy} and SciPy \citep{Scipya,Scipyb}. All the figures in this paper are plotted using the Python matplotlib package \citep{Matplotlib}. This research has made use of NASA's Astrophysics Data System and the arXiv preprint server. We further thank Robert Thompson for developing Caesar, and the yt team for the development and support of yt \citep{yt}.

\end{acknowledgements}




\bibliographystyle{aa}
\bibliography{ref}

\begin{thebibliography}{128}
\expandafter\ifx\csname natexlab\endcsname\relax\def\natexlab#1{#1}\fi

\bibitem[{{Aird} {et~al.}(2019){Aird}, {Coil}, \& {Georgakakis}}]{Aird2019}
{Aird}, J., {Coil}, A.~L., \& {Georgakakis}, A. 2019, \mnras, 484, 4360

\bibitem[{{Aird} {et~al.}(2012){Aird}, {Coil}, {Moustakas}, {Blanton}, {Burles}, {Cool}, {Eisenstein}, {Smith}, {Wong}, \& {Zhu}}]{Aird2012}
{Aird}, J., {Coil}, A.~L., {Moustakas}, J., {et~al.} 2012, \apj, 746, 90

\bibitem[{{Alexander} \& {Natarajan}(2014)}]{Alexander2014}
{Alexander}, T. \& {Natarajan}, P. 2014, Science, 345, 1330

\bibitem[{{Angl{\'e}s-Alc{\'a}zar} {et~al.}(2017){Angl{\'e}s-Alc{\'a}zar}, {Dav{\'e}}, {Faucher-Gigu{\`e}re}, {{\"O}zel}, \& {Hopkins}}]{Angles2017}
{Angl{\'e}s-Alc{\'a}zar}, D., {Dav{\'e}}, R., {Faucher-Gigu{\`e}re}, C.-A., {{\"O}zel}, F., \& {Hopkins}, P.~F. 2017, \mnras, 464, 2840

\bibitem[{{Angl{\'e}s-Alc{\'a}zar} {et~al.}(2013){Angl{\'e}s-Alc{\'a}zar}, {{\"O}zel}, \& {Dav{\'e}}}]{Angles2013}
{Angl{\'e}s-Alc{\'a}zar}, D., {{\"O}zel}, F., \& {Dav{\'e}}, R. 2013, \apj, 770, 5

\bibitem[{{Angl{\'e}s-Alc{\'a}zar} {et~al.}(2015){Angl{\'e}s-Alc{\'a}zar}, {{\"O}zel}, {Dav{\'e}}, {Katz}, {Kollmeier}, \& {Oppenheimer}}]{Angles2015}
{Angl{\'e}s-Alc{\'a}zar}, D., {{\"O}zel}, F., {Dav{\'e}}, R., {et~al.} 2015, \apj, 800, 127

\bibitem[{{Ba{\~n}ados} {et~al.}(2018){Ba{\~n}ados}, {Venemans}, {Mazzucchelli}, {Farina}, {Walter}, {Wang}, {Decarli}, {Stern}, {Fan}, {Davies}, {Hennawi}, {Simcoe}, {Turner}, {Rix}, {Yang}, {Kelson}, {Rudie}, \& {Winters}}]{Banados2018}
{Ba{\~n}ados}, E., {Venemans}, B.~P., {Mazzucchelli}, C., {et~al.} 2018, \nat, 553, 473

\bibitem[{{Baron} \& {M{\'e}nard}(2019)}]{Baron2019}
{Baron}, D. \& {M{\'e}nard}, B. 2019, Monthly Notices of the Royal Astronomical Society, 487, 3404

\bibitem[{{Beckmann} {et~al.}(2017){Beckmann}, {Devriendt}, {Slyz}, {Peirani}, {Richardson}, {Dubois}, {Pichon}, {Chisari}, {Kaviraj}, {Laigle}, \& {Volonteri}}]{Beckmann2017}
{Beckmann}, R.~S., {Devriendt}, J., {Slyz}, A., {et~al.} 2017, \mnras, 472, 949

\bibitem[{{Bhowmick} {et~al.}(2020){Bhowmick}, {Blecha}, \& {Thomas}}]{Bhowmick2020}
{Bhowmick}, A.~K., {Blecha}, L., \& {Thomas}, J. 2020, \apj, 904, 150

\bibitem[{{Bhowmick} {et~al.}(2025){Bhowmick}, {Blecha}, {Torrey}, {Somerville}, {Kelley}, {Weinberger}, {Vogelsberger}, {Hernquist}, {Natarajan}, {Kho}, \& {Di Matteo}}]{Bhowmick2025}
{Bhowmick}, A.~K., {Blecha}, L., {Torrey}, P., {et~al.} 2025, \mnras, 538, 518

\bibitem[{{Bigiel} {et~al.}(2008){Bigiel}, {Leroy}, {Walter}, {Brinks}, {de Blok}, {Madore}, \& {Thornley}}]{Bigiel2008}
{Bigiel}, F., {Leroy}, A., {Walter}, F., {et~al.} 2008, \aj, 136, 2846

\bibitem[{{Birchall} {et~al.}(2020){Birchall}, {Watson}, \& {Aird}}]{Birchall2020}
{Birchall}, K.~L., {Watson}, M.~G., \& {Aird}, J. 2020, Monthly Notices of the Royal Astronomical Society, 492, 2268

\bibitem[{{Blank} {et~al.}(2019){Blank}, {Macci{\`o}}, {Dutton}, \& {Obreja}}]{Blank2019}
{Blank}, M., {Macci{\`o}}, A.~V., {Dutton}, A.~A., \& {Obreja}, A. 2019, \mnras, 487, 5476

\bibitem[{{Bluck} {et~al.}(2020){Bluck}, {Maiolino}, {Piotrowska}, {Trussler}, {Ellison}, {S{\'a}nchez}, {Thorp}, {Teimoorinia}, {Moreno}, \& {Conselice}}]{Bluck2020}
{Bluck}, A. F.~L., {Maiolino}, R., {Piotrowska}, J.~M., {et~al.} 2020, \mnras, 499, 230

\bibitem[{{Boco} {et~al.}(2023){Boco}, {Lapi}, {Shankar}, {Fu}, {Gabrielli}, \& {Sicilia}}]{Boco2023}
{Boco}, L., {Lapi}, A., {Shankar}, F., {et~al.} 2023, The Astrophysical Journal, 954, 97

\bibitem[{{Bondi}(1952)}]{Bondi1952}
{Bondi}, H. 1952, \mnras, 112, 195

\bibitem[{{Bongiorno} {et~al.}(2014){Bongiorno}, {Maiolino}, {Brusa}, {Marconi}, {Piconcelli}, {Lamastra}, {Cano-D{\'\i}az}, {Schulze}, {Magnelli}, {Vignali}, {Fiore}, {Menci}, {Cresci}, {La Franca}, \& {Merloni}}]{Bongiorno2014}
{Bongiorno}, A., {Maiolino}, R., {Brusa}, M., {et~al.} 2014, \mnras, 443, 2077

\bibitem[{{Bongiorno} {et~al.}(2016){Bongiorno}, {Schulze}, {Merloni}, {Zamorani}, {Ilbert}, {La Franca}, {Peng}, {Piconcelli}, {Mainieri}, {Silverman}, {Brusa}, {Fiore}, {Salvato}, \& {Scoville}}]{Bongiorno2016}
{Bongiorno}, A., {Schulze}, A., {Merloni}, A., {et~al.} 2016, \aap, 588, A78

\bibitem[{{Bower} {et~al.}(2017){Bower}, {Schaye}, {Frenk}, {Theuns}, {Schaller}, {Crain}, \& {McAlpine}}]{Bower2017}
{Bower}, R.~G., {Schaye}, J., {Frenk}, C.~S., {et~al.} 2017, \mnras, 465, 32

\bibitem[{{Brown} {et~al.}(2017){Brown}, {Catinella}, {Cortese}, {Lagos}, {Dav{\'e}}, {Kilborn}, {Haynes}, {Giovanelli}, \& {Rafieferantsoa}}]{Brown2017}
{Brown}, T., {Catinella}, B., {Cortese}, L., {et~al.} 2017, \mnras, 466, 1275

\bibitem[{{Chen} {et~al.}(2013){Chen}, {Hickox}, {Alberts}, {Brodwin}, {Jones}, {Murray}, {Alexander}, {Assef}, {Brown}, {Dey}, {Forman}, {Gorjian}, {Goulding}, {Le Floc'h}, {Jannuzi}, {Mullaney}, \& {Pope}}]{Chen2013}
{Chen}, C.-T.~J., {Hickox}, R.~C., {Alberts}, S., {et~al.} 2013, \apj, 773, 3

\bibitem[{{Cisternas} {et~al.}(2011){Cisternas}, {Jahnke}, {Bongiorno}, {Inskip}, {Impey}, {Koekemoer}, {Merloni}, {Salvato}, \& {Trump}}]{Cisternas2011}
{Cisternas}, M., {Jahnke}, K., {Bongiorno}, A., {et~al.} 2011, \apjl, 741, L11

\bibitem[{{Cortese} {et~al.}(2021){Cortese}, {Catinella}, \& {Smith}}]{Cortese2021}
{Cortese}, L., {Catinella}, B., \& {Smith}, R. 2021, \pasa, 38, e035

\bibitem[{{Crain} {et~al.}(2015){Crain}, {Schaye}, {Bower}, {Furlong}, {Schaller}, {Theuns}, {Dalla Vecchia}, {Frenk}, {McCarthy}, {Helly}, {Jenkins}, {Rosas-Guevara}, {White}, \& {Trayford}}]{Crain2015}
{Crain}, R.~A., {Schaye}, J., {Bower}, R.~G., {et~al.} 2015, \mnras, 450, 1937

\bibitem[{{Cui} {et~al.}(2022){Cui}, {Dave}, {Knebe}, {Rasia}, {Gray}, {Pearce}, {Power}, {Yepes}, {Anbajagane}, {Ceverino}, {Contreras-Santos}, {de Andres}, {De Petris}, {Ettori}, {Haggar}, {Li}, {Wang}, {Yang}, {Borgani}, {Dolag}, {Zu}, {Kuchner}, {Ca{\~n}as}, {Ferragamo}, \& {Gianfagna}}]{Cui2022}
{Cui}, W., {Dave}, R., {Knebe}, A., {et~al.} 2022, \mnras, 514, 977

\bibitem[{{Cui} {et~al.}(2021){Cui}, {Dav{\'e}}, {Peacock}, {Angl{\'e}s-Alc{\'a}zar}, \& {Yang}}]{Cui2021}
{Cui}, W., {Dav{\'e}}, R., {Peacock}, J.~A., {Angl{\'e}s-Alc{\'a}zar}, D., \& {Yang}, X. 2021, Nature Astronomy, 5, 1069

\bibitem[{{Cui} {et~al.}(2024){Cui}, {Jennings}, {Dave}, {Babul}, \& {Gozaliasl}}]{Cui2024}
{Cui}, W., {Jennings}, F., {Dave}, R., {Babul}, A., \& {Gozaliasl}, G. 2024, \mnras, 534, 1247

\bibitem[{{Cui} {et~al.}(2018){Cui}, {Knebe}, {Yepes}, {Pearce}, {Power}, {Dave}, {Arth}, {Borgani}, {Dolag}, \& {Elahi}}]{Cui2018b}
{Cui}, W., {Knebe}, A., {Yepes}, G., {et~al.} 2018, \mnras, 480, 2898

\bibitem[{{Dav{\'e}} {et~al.}(2019){Dav{\'e}}, {Angl{\'e}s-Alc{\'a}zar}, {Narayanan}, {Li}, {Rafieferantsoa}, \& {Appleby}}]{Dave2019}
{Dav{\'e}}, R., {Angl{\'e}s-Alc{\'a}zar}, D., {Narayanan}, D., {et~al.} 2019, \mnras, 486, 2827

\bibitem[{{Dav{\'e}} {et~al.}(2016){Dav{\'e}}, {Thompson}, \& {Hopkins}}]{Dave2016}
{Dav{\'e}}, R., {Thompson}, R., \& {Hopkins}, P.~F. 2016, \mnras, 462, 3265

\bibitem[{{Davis} {et~al.}(2018){Davis}, {Graham}, \& {Cameron}}]{Davis2018}
{Davis}, B.~L., {Graham}, A.~W., \& {Cameron}, E. 2018, The Astrophysical Journal, 869, 113

\bibitem[{{Davis} {et~al.}(2019){Davis}, {Graham}, \& {Cameron}}]{Davis2019}
{Davis}, B.~L., {Graham}, A.~W., \& {Cameron}, E. 2019, \apj, 873, 85

\bibitem[{{Di Matteo} {et~al.}(2005){Di Matteo}, {Springel}, \& {Hernquist}}]{Di2005}
{Di Matteo}, T., {Springel}, V., \& {Hernquist}, L. 2005, \nat, 433, 604

\bibitem[{{Ding} {et~al.}(2020){Ding}, {Silverman}, {Treu}, {Schulze}, {Schramm}, {Birrer}, {Park}, {Jahnke}, {Bennert}, {Kartaltepe}, {Koekemoer}, {Malkan}, \& {Sanders}}]{Ding2020}
{Ding}, X., {Silverman}, J., {Treu}, T., {et~al.} 2020, \apj, 888, 37

\bibitem[{{Donnari} {et~al.}(2021){Donnari}, {Pillepich}, {Nelson}, {Marinacci}, {Vogelsberger}, \& {Hernquist}}]{Donnari2021}
{Donnari}, M., {Pillepich}, A., {Nelson}, D., {et~al.} 2021, \mnras, 506, 4760

\bibitem[{{Dubois} {et~al.}(2016){Dubois}, {Peirani}, {Pichon}, {Devriendt}, {Gavazzi}, {Welker}, \& {Volonteri}}]{Dubois2016}
{Dubois}, Y., {Peirani}, S., {Pichon}, C., {et~al.} 2016, \mnras, 463, 3948

\bibitem[{{Graham}(2008)}]{Graham2008}
{Graham}, A.~W. 2008, \apj, 680, 143

\bibitem[{{Graham}(2012)}]{Graham2012}
{Graham}, A.~W. 2012, The Astrophysical Journal, 746, 113

\bibitem[{{Graham} \& {Sahu}(2023)}]{Graham2023}
{Graham}, A.~W. \& {Sahu}, N. 2023, \mnras, 518, 2177

\bibitem[{{Greene} {et~al.}(2020){Greene}, {Strader}, \& {Ho}}]{Greene2020}
{Greene}, J.~E., {Strader}, J., \& {Ho}, L.~C. 2020, \araa, 58, 257

\bibitem[{{G{\"u}ltekin} {et~al.}(2009){G{\"u}ltekin}, {Richstone}, {Gebhardt}, {Lauer}, {Tremaine}, {Aller}, {Bender}, {Dressler}, {Faber}, {Filippenko}, {Green}, {Ho}, {Kormendy}, {Magorrian}, {Pinkney}, \& {Siopis}}]{Gultekin2009}
{G{\"u}ltekin}, K., {Richstone}, D.~O., {Gebhardt}, K., {et~al.} 2009, \apj, 698, 198

\bibitem[{{Habouzit} {et~al.}(2021){Habouzit}, {Li}, {Somerville}, {Genel}, {Pillepich}, {Volonteri}, {Dav{\'e}}, {Rosas-Guevara}, {McAlpine}, {Peirani}, {Hernquist}, {Angl{\'e}s-Alc{\'a}zar}, {Reines}, {Bower}, {Dubois}, {Nelson}, {Pichon}, \& {Vogelsberger}}]{Habouzit2021}
{Habouzit}, M., {Li}, Y., {Somerville}, R.~S., {et~al.} 2021, \mnras, 503, 1940

\bibitem[{{Habouzit} {et~al.}(2022){Habouzit}, {Onoue}, {Ba{\~n}ados}, {Neeleman}, {Angl{\'e}s-Alc{\'a}zar}, {Walter}, {Pillepich}, {Dav{\'e}}, {Jahnke}, \& {Dubois}}]{Habouzit2022}
{Habouzit}, M., {Onoue}, M., {Ba{\~n}ados}, E., {et~al.} 2022, \mnras, 511, 3751

\bibitem[{{Haemmerl{\'e}} {et~al.}(2020){Haemmerl{\'e}}, {Mayer}, {Klessen}, {Hosokawa}, {Madau}, \& {Bromm}}]{Haemmerle2020}
{Haemmerl{\'e}}, L., {Mayer}, L., {Klessen}, R.~S., {et~al.} 2020, Space Science Reviews, 216, 48

\bibitem[{{Harikane} {et~al.}(2023){Harikane}, {Zhang}, {Nakajima}, {Ouchi}, {Isobe}, {Ono}, {Hatano}, {Xu}, \& {Umeda}}]{Harikane2023}
{Harikane}, Y., {Zhang}, Y., {Nakajima}, K., {et~al.} 2023, \apj, 959, 39

\bibitem[{{H{\"a}ring} \& {Rix}(2004)}]{Haring2004}
{H{\"a}ring}, N. \& {Rix}, H.-W. 2004, \apjl, 604, L89

\bibitem[{{Heckman} \& {Best}(2014)}]{Heckman2014}
{Heckman}, T.~M. \& {Best}, P.~N. 2014, \araa, 52, 589

\bibitem[{{Hirschmann} {et~al.}(2010){Hirschmann}, {Khochfar}, {Burkert}, {Naab}, {Genel}, \& {Somerville}}]{Hirschmann2010}
{Hirschmann}, M., {Khochfar}, S., {Burkert}, A., {et~al.} 2010, \mnras, 407, 1016

\bibitem[{{Hirschmann} {et~al.}(2015){Hirschmann}, {Naab}, {Ostriker}, {Forbes}, {Duc}, {Dav{\'e}}, {Oser}, \& {Karabal}}]{Hirschmann2015}
{Hirschmann}, M., {Naab}, T., {Ostriker}, J.~P., {et~al.} 2015, \mnras, 449, 528

\bibitem[{{Hopkins}(2015)}]{Hopkins2015}
{Hopkins}, P.~F. 2015, \mnras, 450, 53

\bibitem[{{Hopkins} \& {Quataert}(2011)}]{Hopkins2011}
{Hopkins}, P.~F. \& {Quataert}, E. 2011, \mnras, 415, 1027

\bibitem[{{Hu} {et~al.}(2022){Hu}, {Inayoshi}, {Haiman}, {Li}, {Quataert}, \& {Kuiper}}]{Hu2022}
{Hu}, H., {Inayoshi}, K., {Haiman}, Z., {et~al.} 2022, \apj, 935, 140

\bibitem[{Hunter(2007)}]{Matplotlib}
Hunter, J.~D. 2007, Computing In Science \& Engineering, 9, 90

\bibitem[{{Kennicutt}(1998)}]{Kennicutt1998}
{Kennicutt}, Robert~C., J. 1998, \apj, 498, 541

\bibitem[{{Kennicutt} {et~al.}(2007){Kennicutt}, {Calzetti}, {Walter}, {Helou}, {Hollenbach}, {Armus}, {Bendo}, {Dale}, {Draine}, {Engelbracht}, {Gordon}, {Prescott}, {Regan}, {Thornley}, {Bot}, {Brinks}, {de Blok}, {de Mello}, {Meyer}, {Moustakas}, {Murphy}, {Sheth}, \& {Smith}}]{Kennicutt2007}
{Kennicutt}, Jr., R.~C., {Calzetti}, D., {Walter}, F., {et~al.} 2007, \apj, 671, 333

\bibitem[{{Kim} {et~al.}(2008){Kim}, {Ho}, {Peng}, {Barth}, {Im}, {Martini}, \& {Nelson}}]{Kim2008}
{Kim}, M., {Ho}, L.~C., {Peng}, C.~Y., {et~al.} 2008, \apj, 687, 767

\bibitem[{{Kocevski} {et~al.}(2023){Kocevski}, {Onoue}, {Inayoshi}, {Trump}, {Arrabal Haro}, {Grazian}, {Dickinson}, {Finkelstein}, {Kartaltepe}, {Hirschmann}, {Aird}, {Holwerda}, {Fujimoto}, {Juneau}, {Amor{\'\i}n}, {Backhaus}, {Bagley}, {Barro}, {Bell}, {Bisigello}, {Calabr{\`o}}, {Cleri}, {Cooper}, {Ding}, {Grogin}, {Ho}, {Hutchison}, {Inoue}, {Jiang}, {Jones}, {Koekemoer}, {Li}, {Li}, {McGrath}, {Molina}, {Papovich}, {P{\'e}rez-Gonz{\'a}lez}, {Pirzkal}, {Wilkins}, {Yang}, \& {Yung}}]{Kocevski2023}
{Kocevski}, D.~D., {Onoue}, M., {Inayoshi}, K., {et~al.} 2023, \apjl, 954, L4

\bibitem[{{Kokorev} {et~al.}(2023){Kokorev}, {Fujimoto}, {Labbe}, {Greene}, {Bezanson}, {Dayal}, {Nelson}, {Atek}, {Brammer}, {Caputi}, {Chemerynska}, {Cutler}, {Feldmann}, {Fudamoto}, {Furtak}, {Goulding}, {de Graaff}, {Leja}, {Marchesini}, {Miller}, {Nanayakkara}, {Oesch}, {Pan}, {Price}, {Setton}, {Smit}, {Stefanon}, {Wang}, {Weaver}, {Whitaker}, {Williams}, \& {Zitrin}}]{Kokorev2023}
{Kokorev}, V., {Fujimoto}, S., {Labbe}, I., {et~al.} 2023, \apjl, 957, L7

\bibitem[{{Kormendy} \& {Ho}(2013)}]{Kormendy2013}
{Kormendy}, J. \& {Ho}, L.~C. 2013, \araa, 51, 511

\bibitem[{{Kormendy} \& {Richstone}(1995)}]{Kormendy1995}
{Kormendy}, J. \& {Richstone}, D. 1995, \araa, 33, 581

\bibitem[{{Krumholz} \& {Gnedin}(2011)}]{Krumholz2011}
{Krumholz}, M.~R. \& {Gnedin}, N.~Y. 2011, \apj, 729, 36

\bibitem[{{Larson} {et~al.}(2023){Larson}, {Finkelstein}, {Kocevski}, {Hutchison}, {Trump}, {Arrabal Haro}, {Bromm}, {Cleri}, {Dickinson}, {Fujimoto}, {Kartaltepe}, {Koekemoer}, {Papovich}, {Pirzkal}, {Tacchella}, {Zavala}, {Bagley}, {Behroozi}, {Champagne}, {Cole}, {Jung}, {Morales}, {Yang}, {Zhang}, {Zitrin}, {Amor{\'\i}n}, {Burgarella}, {Casey}, {Ch{\'a}vez Ortiz}, {Cox}, {Chworowsky}, {Fontana}, {Gawiser}, {Grazian}, {Grogin}, {Harish}, {Hathi}, {Hirschmann}, {Holwerda}, {Juneau}, {Leung}, {Lucas}, {McGrath}, {P{\'e}rez-Gonz{\'a}lez}, {Rigby}, {Seill{\'e}}, {Simons}, {de La Vega}, {Weiner}, {Wilkins}, {Yung}, \& {Ceers Team}}]{Larson2023}
{Larson}, R.~L., {Finkelstein}, S.~L., {Kocevski}, D.~D., {et~al.} 2023, \apjl, 953, L29

\bibitem[{{Li} {et~al.}(2024){Li}, {Chen}, {Wang}, \& {Mo}}]{Li2024}
{Li}, H., {Chen}, Y., {Wang}, H., \& {Mo}, H. 2024, MNRAS, subm., arXiv:2409.06208

\bibitem[{{Li} {et~al.}(2025){Li}, {Silverman}, {Shen}, {Volonteri}, {Jahnke}, {Zhuang}, {Scoggins}, {Ding}, {Harikane}, {Onoue}, \& {Tanaka}}]{Li2025}
{Li}, J., {Silverman}, J.~D., {Shen}, Y., {et~al.} 2025, \apj, 981, 19

\bibitem[{{Li} {et~al.}(2023){Li}, {Shen}, {Ho}, {Brandt}, {Grier}, {Hall}, {Homayouni}, {Koekemoer}, {Schneider}, \& {Trump}}]{Li2023}
{Li}, J. I.~H., {Shen}, Y., {Ho}, L.~C., {et~al.} 2023, The Astrophysical Journal, 954, 173

\bibitem[{{Li} {et~al.}(2020){Li}, {Habouzit}, {Genel}, {Somerville}, {Terrazas}, {Bell}, {Pillepich}, {Nelson}, {Weinberger}, {Rodriguez-Gomez}, {Ma}, {Pakmor}, {Hernquist}, \& {Vogelsberger}}]{Li2020}
{Li}, Y., {Habouzit}, M., {Genel}, S., {et~al.} 2020, \apj, 895, 102

\bibitem[{{Liu} {et~al.}(2025){Liu}, {Guo}, {Wang}, {Xu}, {Lu}, {Cui}, \& {Dav{\'e}}}]{Liu2025}
{Liu}, K., {Guo}, H., {Wang}, S., {et~al.} 2025, \aap, 693, A48

\bibitem[{{Ma} {et~al.}(2022){Ma}, {Liu}, {Guo}, {Cui}, {Jones}, {Wang}, {Zhang}, \& {Dav{\'e}}}]{Ma2022}
{Ma}, W., {Liu}, K., {Guo}, H., {et~al.} 2022, \apj, 941, 205

\bibitem[{{Magorrian} {et~al.}(1998){Magorrian}, {Tremaine}, {Richstone}, {Bender}, {Bower}, {Dressler}, {Faber}, {Gebhardt}, {Green}, {Grillmair}, {Kormendy}, \& {Lauer}}]{Magorrian1998}
{Magorrian}, J., {Tremaine}, S., {Richstone}, D., {et~al.} 1998, \aj, 115, 2285

\bibitem[{{Maiolino} {et~al.}(2024{\natexlab{a}}){Maiolino}, {Scholtz}, {Curtis-Lake}, {Carniani}, {Baker}, {de Graaff}, {Tacchella}, {{\"U}bler}, {D'Eugenio}, {Witstok}, {Curti}, {Arribas}, {Bunker}, {Charlot}, {Chevallard}, {Eisenstein}, {Egami}, {Ji}, {Jones}, {Lyu}, {Rawle}, {Robertson}, {Rujopakarn}, {Perna}, {Sun}, {Venturi}, {Williams}, \& {Willott}}]{Maiolino2024b}
{Maiolino}, R., {Scholtz}, J., {Curtis-Lake}, E., {et~al.} 2024{\natexlab{a}}, \aap, 691, A145

\bibitem[{{Maiolino} {et~al.}(2024{\natexlab{b}}){Maiolino}, {Scholtz}, {Witstok}, {Carniani}, {D'Eugenio}, {de Graaff}, {{\"U}bler}, {Tacchella}, {Curtis-Lake}, {Arribas}, {Bunker}, {Charlot}, {Chevallard}, {Curti}, {Looser}, {Maseda}, {Rawle}, {Rodr{\'\i}guez del Pino}, {Willott}, {Egami}, {Eisenstein}, {Hainline}, {Robertson}, {Williams}, {Willmer}, {Baker}, {Boyett}, {DeCoursey}, {Fabian}, {Helton}, {Ji}, {Jones}, {Kumari}, {Laporte}, {Nelson}, {Perna}, {Sandles}, {Shivaei}, \& {Sun}}]{Maiolino2024a}
{Maiolino}, R., {Scholtz}, J., {Witstok}, J., {et~al.} 2024{\natexlab{b}}, \nat, 627, 59

\bibitem[{{Manzano-King} {et~al.}(2019){Manzano-King}, {Canalizo}, \& {Sales}}]{Manzano-King2019}
{Manzano-King}, C.~M., {Canalizo}, G., \& {Sales}, L.~V. 2019, The Astrophysical Journal, 884, 54

\bibitem[{{Marasco} {et~al.}(2021){Marasco}, {Cresci}, {Posti}, {Fraternali}, {Mannucci}, {Marconi}, {Belfiore}, \& {Fall}}]{Marasco2021}
{Marasco}, A., {Cresci}, G., {Posti}, L., {et~al.} 2021, \mnras, 507, 4274

\bibitem[{{Mart{\'\i}n-Navarro} {et~al.}(2018){Mart{\'\i}n-Navarro}, {Brodie}, {Romanowsky}, {Ruiz-Lara}, \& {van de Ven}}]{Martin-Navarro2018a}
{Mart{\'\i}n-Navarro}, I., {Brodie}, J.~P., {Romanowsky}, A.~J., {Ruiz-Lara}, T., \& {van de Ven}, G. 2018, \nat, 553, 307

\bibitem[{{Mart{\'\i}n-Navarro} \& {Mezcua}(2018)}]{Martin-Navarro2018b}
{Mart{\'\i}n-Navarro}, I. \& {Mezcua}, M. 2018, \apjl, 855, L20

\bibitem[{{McAlpine} {et~al.}(2018){McAlpine}, {Bower}, {Rosario}, {Crain}, {Schaye}, \& {Theuns}}]{McAlpine2018}
{McAlpine}, S., {Bower}, R.~G., {Rosario}, D.~J., {et~al.} 2018, \mnras, 481, 3118

\bibitem[{{McCarthy} {et~al.}(2011){McCarthy}, {Schaye}, {Bower}, {Ponman}, {Booth}, {Dalla Vecchia}, \& {Springel}}]{McCarthy2011}
{McCarthy}, I.~G., {Schaye}, J., {Bower}, R.~G., {et~al.} 2011, \mnras, 412, 1965

\bibitem[{{McConnell} \& {Ma}(2013)}]{McConnell2013}
{McConnell}, N.~J. \& {Ma}, C.-P. 2013, \apj, 764, 184

\bibitem[{{McLure} \& {Dunlop}(2002)}]{McLure2002}
{McLure}, R.~J. \& {Dunlop}, J.~S. 2002, \mnras, 331, 795

\bibitem[{{Merritt} \& {Ferrarese}(2001)}]{Merritt2001}
{Merritt}, D. \& {Ferrarese}, L. 2001, \apj, 547, 140

\bibitem[{{Mezcua} {et~al.}(2018{\natexlab{a}}){Mezcua}, {Civano}, {Marchesi}, {Suh}, {Fabbiano}, \& {Volonteri}}]{Mezcua2018a}
{Mezcua}, M., {Civano}, F., {Marchesi}, S., {et~al.} 2018{\natexlab{a}}, \mnras, 478, 2576

\bibitem[{{Mezcua} \& {Dom{\'\i}nguez S{\'a}nchez}(2024)}]{Mezcua2024}
{Mezcua}, M. \& {Dom{\'\i}nguez S{\'a}nchez}, H. 2024, Monthly Notices of the Royal Astronomical Society, 528, 5252

\bibitem[{{Mezcua} {et~al.}(2018{\natexlab{b}}){Mezcua}, {Kim}, {Ho}, \& {Lonsdale}}]{Mezcua2018}
{Mezcua}, M., {Kim}, M., {Ho}, L.~C., \& {Lonsdale}, C.~J. 2018{\natexlab{b}}, \mnras, 480, L74

\bibitem[{Millman \& Aivazis(2011)}]{Scipyb}
Millman, K.~J. \& Aivazis, M. 2011, Computing in Science and Engineering, 13, 9

\bibitem[{{Mo} {et~al.}(2024){Mo}, {Chen}, \& {Wang}}]{Mo2024}
{Mo}, H., {Chen}, Y., \& {Wang}, H. 2024, \mnras, 532, 3808

\bibitem[{{Molina} {et~al.}(2024){Molina}, {Ho}, \& {Knudsen}}]{Molina2024}
{Molina}, J., {Ho}, L.~C., \& {Knudsen}, K.~K. 2024, \aap, 691, A114

\bibitem[{{Mountrichas}(2023)}]{Mountrichas2023}
{Mountrichas}, G. 2023, \aap, 672, A98

\bibitem[{{Nelson} {et~al.}(2015){Nelson}, {Pillepich}, {Genel}, {Vogelsberger}, {Springel}, {Torrey}, {Rodriguez-Gomez}, {Sijacki}, {Snyder}, {Griffen}, {Marinacci}, {Blecha}, {Sales}, {Xu}, \& {Hernquist}}]{Nelson2015}
{Nelson}, D., {Pillepich}, A., {Genel}, S., {et~al.} 2015, Astronomy and Computing, 13, 12

\bibitem[{{Newville} {et~al.}(2014){Newville}, {Stensitzki}, {Allen}, \& {Ingargiola}}]{LMFIT}
{Newville}, M., {Stensitzki}, T., {Allen}, D.~B., \& {Ingargiola}, A. 2014, {LMFIT: Non-Linear Least-Square Minimization and Curve-Fitting for Python}

\bibitem[{Oliphant(2007)}]{Scipya}
Oliphant, T.~E. 2007, Computing in Science and Engg., 9, 10

\bibitem[{{Pacucci} {et~al.}(2023){Pacucci}, {Nguyen}, {Carniani}, {Maiolino}, \& {Fan}}]{Pacucci2023}
{Pacucci}, F., {Nguyen}, B., {Carniani}, S., {Maiolino}, R., \& {Fan}, X. 2023, \apjl, 957, L3

\bibitem[{{Page} {et~al.}(2012){Page}, {Symeonidis}, {Vieira}, {Altieri}, {Amblard}, {Arumugam}, {Aussel}, {Babbedge}, {Blain}, {Bock}, {Boselli}, {Buat}, {Castro-Rodr{\'\i}guez}, {Cava}, {Chanial}, {Clements}, {Conley}, {Conversi}, {Cooray}, {Dowell}, {Dubois}, {Dunlop}, {Dwek}, {Dye}, {Eales}, {Elbaz}, {Farrah}, {Fox}, {Franceschini}, {Gear}, {Glenn}, {Griffin}, {Halpern}, {Hatziminaoglou}, {Ibar}, {Isaak}, {Ivison}, {Lagache}, {Levenson}, {Lu}, {Madden}, {Maffei}, {Mainetti}, {Marchetti}, {Nguyen}, {O'Halloran}, {Oliver}, {Omont}, {Panuzzo}, {Papageorgiou}, {Pearson}, {P{\'e}rez-Fournon}, {Pohlen}, {Rawlings}, {Rigopoulou}, {Riguccini}, {Rizzo}, {Rodighiero}, {Roseboom}, {Rowan-Robinson}, {Portal}, {Schulz}, {Scott}, {Seymour}, {Shupe}, {Smith}, {Stevens}, {Trichas}, {Tugwell}, {Vaccari}, {Valtchanov}, {Viero}, {Vigroux}, {Wang}, {Ward}, {Wright}, {Xu}, \& {Zemcov}}]{Page2012}
{Page}, M.~J., {Symeonidis}, M., {Vieira}, J.~D., {et~al.} 2012, \nat, 485, 213

\bibitem[{P\'erez \& Granger(2007)}]{ipython}
P\'erez, F. \& Granger, B.~E. 2007, Computing in Science and Engineering, 9, 21

\bibitem[{{Pillepich} {et~al.}(2018){Pillepich}, {Springel}, {Nelson}, {Genel}, {Naiman}, {Pakmor}, {Hernquist}, {Torrey}, {Vogelsberger}, {Weinberger}, \& {Marinacci}}]{Pillepich2018}
{Pillepich}, A., {Springel}, V., {Nelson}, D., {et~al.} 2018, \mnras, 473, 4077

\bibitem[{{Planck Collaboration} {et~al.}(2016){Planck Collaboration}, {Ade}, {Aghanim}, {Arnaud}, {Ashdown}, {Aumont}, {Baccigalupi}, {Banday}, {Barreiro}, {Bartlett}, {Bartolo}, {Battaner}, {Battye}, {Benabed}, {Beno{\^\i}t}, {Benoit-L{\'e}vy}, {Bernard}, {Bersanelli}, {Bielewicz}, {Bock}, {Bonaldi}, {Bonavera}, {Bond}, {Borrill}, {Bouchet}, {Boulanger}, {Bucher}, {Burigana}, {Butler}, {Calabrese}, {Cardoso}, {Catalano}, {Challinor}, {Chamballu}, {Chary}, {Chiang}, {Chluba}, {Christensen}, {Church}, {Clements}, {Colombi}, {Colombo}, {Combet}, {Coulais}, {Crill}, {Curto}, {Cuttaia}, {Danese}, {Davies}, {Davis}, {de Bernardis}, {de Rosa}, {de Zotti}, {Delabrouille}, {D{\'e}sert}, {Di Valentino}, {Dickinson}, {Diego}, {Dolag}, {Dole}, {Donzelli}, {Dor{\'e}}, {Douspis}, {Ducout}, {Dunkley}, {Dupac}, {Efstathiou}, {Elsner}, {En{\ss}lin}, {Eriksen}, {Farhang}, {Fergusson}, {Finelli}, {Forni}, {Frailis}, {Fraisse}, {Franceschi}, {Frejsel}, {Galeotta}, {Galli}, {Ganga}, {Gauthier}, {Gerbino}, {Ghosh}, {Giard},
  {Giraud-H{\'e}raud}, {Giusarma}, {Gjerl{\o}w}, {Gonz{\'a}lez-Nuevo}, {G{\'o}rski}, {Gratton}, {Gregorio}, {Gruppuso}, {Gudmundsson}, {Hamann}, {Hansen}, {Hanson}, {Harrison}, {Helou}, {Henrot-Versill{\'e}}, {Hern{\'a}ndez-Monteagudo}, {Herranz}, {Hildebrandt}, {Hivon}, {Hobson}, {Holmes}, {Hornstrup}, {Hovest}, {Huang}, {Huffenberger}, {Hurier}, {Jaffe}, {Jaffe}, {Jones}, {Juvela}, {Keih{\"a}nen}, {Keskitalo}, {Kisner}, {Kneissl}, {Knoche}, {Knox}, {Kunz}, {Kurki-Suonio}, {Lagache}, {L{\"a}hteenm{\"a}ki}, {Lamarre}, {Lasenby}, {Lattanzi}, {Lawrence}, {Leahy}, {Leonardi}, {Lesgourgues}, {Levrier}, {Lewis}, {Liguori}, {Lilje}, {Linden-V{\o}rnle}, {L{\'o}pez-Caniego}, {Lubin}, {Mac{\'\i}as-P{\'e}rez}, {Maggio}, {Maino}, {Mandolesi}, {Mangilli}, {Marchini}, {Maris}, {Martin}, {Martinelli}, {Mart{\'\i}nez-Gonz{\'a}lez}, {Masi}, {Matarrese}, {McGehee}, {Meinhold}, {Melchiorri}, {Melin}, {Mendes}, {Mennella}, {Migliaccio}, {Millea}, {Mitra}, {Miville-Desch{\^e}nes}, {Moneti}, {Montier}, {Morgante}, {Mortlock},
  {Moss}, {Munshi}, {Murphy}, {Naselsky}, {Nati}, {Natoli}, {Netterfield}, {N{\o}rgaard-Nielsen}, {Noviello}, {Novikov}, {Novikov}, {Oxborrow}, {Paci}, {Pagano}, {Pajot}, {Paladini}, {Paoletti}, {Partridge}, {Pasian}, {Patanchon}, {Pearson}, {Perdereau}, {Perotto}, {Perrotta}, {Pettorino}, {Piacentini}, {Piat}, {Pierpaoli}, {Pietrobon}, {Plaszczynski}, {Pointecouteau}, {Polenta}, {Popa}, {Pratt}, {Pr{\'e}zeau}, {Prunet}, {Puget}, {Rachen}, {Reach}, {Rebolo}, {Reinecke}, {Remazeilles}, {Renault}, {Renzi}, {Ristorcelli}, {Rocha}, {Rosset}, {Rossetti}, {Roudier}, {Rouill{\'e} d'Orfeuil}, {Rowan-Robinson}, {Rubi{\~n}o-Mart{\'\i}n}, {Rusholme}, {Said}, {Salvatelli}, {Salvati}, {Sandri}, {Santos}, {Savelainen}, {Savini}, {Scott}, {Seiffert}, {Serra}, {Shellard}, {Spencer}, {Spinelli}, {Stolyarov}, {Stompor}, {Sudiwala}, {Sunyaev}, {Sutton}, {Suur-Uski}, {Sygnet}, {Tauber}, {Terenzi}, {Toffolatti}, {Tomasi}, {Tristram}, {Trombetti}, {Tucci}, {Tuovinen}, {T{\"u}rler}, {Umana}, {Valenziano}, {Valiviita}, {Van Tent},
  {Vielva}, {Villa}, {Wade}, {Wandelt}, {Wehus}, {White}, {White}, {Wilkinson}, {Yvon}, {Zacchei}, \& {Zonca}}]{Planck2016}
{Planck Collaboration}, {Ade}, P.~A.~R., {Aghanim}, N., {et~al.} 2016, \aap, 594, A13

\bibitem[{{Rahmati} {et~al.}(2013){Rahmati}, {Pawlik}, {Rai{\v{c}}evi{\'c}}, \& {Schaye}}]{Rahmati2013}
{Rahmati}, A., {Pawlik}, A.~H., {Rai{\v{c}}evi{\'c}}, M., \& {Schaye}, J. 2013, \mnras, 430, 2427

\bibitem[{{Reines} {et~al.}(2020){Reines}, {Condon}, {Darling}, \& {Greene}}]{Reines2020}
{Reines}, A.~E., {Condon}, J.~J., {Darling}, J., \& {Greene}, J.~E. 2020, \apj, 888, 36

\bibitem[{{Reines} \& {Volonteri}(2015)}]{Reines2015}
{Reines}, A.~E. \& {Volonteri}, M. 2015, \apj, 813, 82

\bibitem[{{Sani} {et~al.}(2011){Sani}, {Marconi}, {Hunt}, \& {Risaliti}}]{Sani2011}
{Sani}, E., {Marconi}, A., {Hunt}, L.~K., \& {Risaliti}, G. 2011, \mnras, 413, 1479

\bibitem[{{Schaye} {et~al.}(2015){Schaye}, {Crain}, {Bower}, {Furlong}, {Schaller}, {Theuns}, {Dalla Vecchia}, {Frenk}, {McCarthy}, {Helly}, {Jenkins}, {Rosas-Guevara}, {White}, {Baes}, {Booth}, {Camps}, {Navarro}, {Qu}, {Rahmati}, {Sawala}, {Thomas}, \& {Trayford}}]{Schaye2015}
{Schaye}, J., {Crain}, R.~A., {Bower}, R.~G., {et~al.} 2015, \mnras, 446, 521

\bibitem[{{Schmidt}(1959)}]{Schmidt1959}
{Schmidt}, M. 1959, \apj, 129, 243

\bibitem[{{Shankar} {et~al.}(2025){Shankar}, {Bernardi}, {Roberts}, {Arana-Catania}, {Grubenmann}, {Habouzit}, {Smith}, {Marsden}, {Varadarajan}, {Tetilla}, {Angl{\'e}s-Alc{\'a}zar}, {Boco}, {Farrah}, {Fu}, {Haniewicz}, {Lapi}, {Lovell}, {Menci}, {Powell}, \& {Ricci}}]{Shankar2025}
{Shankar}, F., {Bernardi}, M., {Roberts}, D., {et~al.} 2025, \mnras [\eprint[arXiv]{2505.02920}]

\bibitem[{{Shi} {et~al.}(2022){Shi}, {Peng}, {Diemer}, {Stevens}, {Pillepich}, {Renzini}, {Dou}, {Gao}, {Gu}, {Ho}, {Kong}, {Lagos}, {Li}, {Li}, {Maiolino}, {Mannucci}, {Xie}, \& {Zhang}}]{Shi2022}
{Shi}, J., {Peng}, Y., {Diemer}, B., {et~al.} 2022, \apj, 927, 189

\bibitem[{{Stone} {et~al.}(2024){Stone}, {Lyu}, {Rieke}, {Alberts}, \& {Hainline}}]{Stone2024}
{Stone}, M.~A., {Lyu}, J., {Rieke}, G.~H., {Alberts}, S., \& {Hainline}, K.~N. 2024, \apj, 964, 90

\bibitem[{{Sun} {et~al.}(2015){Sun}, {Trump}, {Brandt}, {Luo}, {Alexander}, {Jahnke}, {Rosario}, {Wang}, \& {Xue}}]{Sun2015}
{Sun}, M., {Trump}, J.~R., {Brandt}, W.~N., {et~al.} 2015, \apj, 802, 14

\bibitem[{{Terrazas} {et~al.}(2024){Terrazas}, {Aird}, \& {Coil}}]{Terrazas2024}
{Terrazas}, B.~A., {Aird}, J., \& {Coil}, A.~L. 2024, APJ, subm., arXiv:2411.08838

\bibitem[{{Terrazas} {et~al.}(2016){Terrazas}, {Bell}, {Henriques}, {White}, {Cattaneo}, \& {Woo}}]{Terrazas2016}
{Terrazas}, B.~A., {Bell}, E.~F., {Henriques}, B. M.~B., {et~al.} 2016, \apjl, 830, L12

\bibitem[{{Terrazas} {et~al.}(2020){Terrazas}, {Bell}, {Pillepich}, {Nelson}, {Somerville}, {Genel}, {Weinberger}, {Habouzit}, {Li}, {Hernquist}, \& {Vogelsberger}}]{Terrazas2020}
{Terrazas}, B.~A., {Bell}, E.~F., {Pillepich}, A., {et~al.} 2020, \mnras, 493, 1888

\bibitem[{{Thomas} {et~al.}(2019){Thomas}, {Dav{\'e}}, {Angl{\'e}s-Alc{\'a}zar}, \& {Jarvis}}]{Thomas2019}
{Thomas}, N., {Dav{\'e}}, R., {Angl{\'e}s-Alc{\'a}zar}, D., \& {Jarvis}, M. 2019, \mnras, 1662

\bibitem[{{Turk} {et~al.}(2011){Turk}, {Smith}, {Oishi}, {Skory}, {Skillman}, {Abel}, \& {Norman}}]{yt}
{Turk}, M.~J., {Smith}, B.~D., {Oishi}, J.~S., {et~al.} 2011, \apjs, 192, 9

\bibitem[{{{\"U}bler} {et~al.}(2023){{\"U}bler}, {Maiolino}, {Curtis-Lake}, {P{\'e}rez-Gonz{\'a}lez}, {Curti}, {Perna}, {Arribas}, {Charlot}, {Marshall}, {D'Eugenio}, {Scholtz}, {Bunker}, {Carniani}, {Ferruit}, {Jakobsen}, {Rix}, {Rodr{\'\i}guez Del Pino}, {Willott}, {Boeker}, {Cresci}, {Jones}, {Kumari}, \& {Rawle}}]{Ubler2023}
{{\"U}bler}, H., {Maiolino}, R., {Curtis-Lake}, E., {et~al.} 2023, \aap, 677, A145

\bibitem[{van~der Walt {et~al.}(2011)van~der Walt, Colbert, \& Varoquaux}]{NumPy}
van~der Walt, S., Colbert, S.~C., \& Varoquaux, G. 2011, CoRR, abs/1102.1523

\bibitem[{{Vogelsberger} {et~al.}(2014){Vogelsberger}, {Genel}, {Springel}, {Torrey}, {Sijacki}, {Xu}, {Snyder}, {Nelson}, \& {Hernquist}}]{Vogelsberger2014}
{Vogelsberger}, M., {Genel}, S., {Springel}, V., {et~al.} 2014, \mnras, 444, 1518

\bibitem[{{Volonteri} {et~al.}(2021){Volonteri}, {Habouzit}, \& {Colpi}}]{Volonteri2021}
{Volonteri}, M., {Habouzit}, M., \& {Colpi}, M. 2021, Nature Reviews Physics, 3, 732

\bibitem[{{Wang} {et~al.}(2019){Wang}, {Taylor}, {Federrath}, \& {Kobayashi}}]{Wang2019}
{Wang}, E.~X., {Taylor}, P., {Federrath}, C., \& {Kobayashi}, C. 2019, \mnras, 483, 4640

\bibitem[{{Wang} {et~al.}(2021{\natexlab{a}}){Wang}, {Yang}, {Fan}, {Hennawi}, {Barth}, {Banados}, {Bian}, {Boutsia}, {Connor}, {Davies}, {Decarli}, {Eilers}, {Farina}, {Green}, {Jiang}, {Li}, {Mazzucchelli}, {Nanni}, {Schindler}, {Venemans}, {Walter}, {Wu}, \& {Yue}}]{Wang2021}
{Wang}, F., {Yang}, J., {Fan}, X., {et~al.} 2021{\natexlab{a}}, \apjl, 907, L1

\bibitem[{{Wang} {et~al.}(2021{\natexlab{b}}){Wang}, {Staveley-Smith}, {Westmeier}, {Catinella}, {Shao}, {Reynolds}, {For}, {Lee}, {Liang}, {Wang}, {Elagali}, {D{\'e}nes}, {Kleiner}, {Koribalski}, {Lee-Waddell}, {Oh}, {Rhee}, {Serra}, {Spekkens}, {Wong}, {Bekki}, {Bigiel}, {Courtois}, {Hess}, {Holwerda}, {McQuinn}, {Pandey-Pommier}, {van der Hulst}, \& {Verdes-Montenegro}}]{Wang2021a}
{Wang}, J., {Staveley-Smith}, L., {Westmeier}, T., {et~al.} 2021{\natexlab{b}}, \apj, 915, 70

\bibitem[{{Wang} {et~al.}(2015){Wang}, {Dutton}, {Stinson}, {Macci{\`o}}, {Penzo}, {Kang}, {Keller}, \& {Wadsley}}]{Wang2015}
{Wang}, L., {Dutton}, A.~A., {Stinson}, G.~S., {et~al.} 2015, \mnras, 454, 83

\bibitem[{{Wang} {et~al.}(2022){Wang}, {Wang}, {For}, {Lee}, {Reynolds}, {Lin}, {Staveley-Smith}, {Shao}, {Wong}, {Catinella}, {Serra}, {Verdes-Montenegro}, {Westmeier}, {Lee-Waddell}, {Koribalski}, {Murugeshan}, {Elagali}, {Kleiner}, {Rhee}, {Bigiel}, {Bosma}, {Holwerda}, {Oh}, \& {Spekkens}}]{Wang2022}
{Wang}, S., {Wang}, J., {For}, B.-Q., {et~al.} 2022, \apj, 927, 66

\bibitem[{{Weinberger} {et~al.}(2017){Weinberger}, {Springel}, {Hernquist}, {Pillepich}, {Marinacci}, {Pakmor}, {Nelson}, {Genel}, {Vogelsberger}, {Naiman}, \& {Torrey}}]{Weinberger2017}
{Weinberger}, R., {Springel}, V., {Hernquist}, L., {et~al.} 2017, \mnras, 465, 3291

\bibitem[{{Wise}(2019)}]{Wise2019}
{Wise}, J.~H. 2019, in Formation of the First Black Holes, ed. M.~{Latif} \& D.~{Schleicher}, 177--194

\bibitem[{{Wong} \& {Blitz}(2002)}]{Wong2002}
{Wong}, T. \& {Blitz}, L. 2002, \apj, 569, 157

\bibitem[{{Yang} {et~al.}(2023){Yang}, {Caputi}, {Papovich}, {Arrabal Haro}, {Bagley}, {Behroozi}, {Bell}, {Bisigello}, {Buat}, {Burgarella}, {Cheng}, {Cleri}, {Dav{\'e}}, {Dickinson}, {Elbaz}, {Ferguson}, {Finkelstein}, {Grogin}, {Hathi}, {Hirschmann}, {Holwerda}, {Huertas-Company}, {Hutchison}, {Iani}, {Kartaltepe}, {Kirkpatrick}, {Kocevski}, {Koekemoer}, {Kokorev}, {Larson}, {Lucas}, {P{\'e}rez-Gonz{\'a}lez}, {Rinaldi}, {Shen}, {Trump}, {de la Vega}, {Yung}, \& {Zavala}}]{Yang2023}
{Yang}, G., {Caputi}, K.~I., {Papovich}, C., {et~al.} 2023, \apjl, 950, L5

\bibitem[{{Yang} {et~al.}(2024){Yang}, {Dav{\'e}}, {Cui}, {Cai}, {Peacock}, \& {Sorini}}]{Yang2024}
{Yang}, T., {Dav{\'e}}, R., {Cui}, W., {et~al.} 2024, \mnras, 527, 1612

\bibitem[{{Zhu} \& {Springel}(2025)}]{Zhu2025}
{Zhu}, B. \& {Springel}, V. 2025, MNRAS, subm., arXiv:2502.06203

\bibitem[{{Zhuang} \& {Ho}(2023)}]{Zhuang2023}
{Zhuang}, M.-Y. \& {Ho}, L.~C. 2023, Nature Astronomy, 7, 1376

\bibitem[{{Zou} {et~al.}(2024){Zou}, {Brandt}, {Gallo}, {Luo}, {Ni}, {Xue}, \& {Yu}}]{Zou2024}
{Zou}, F., {Brandt}, W.~N., {Gallo}, E., {et~al.} 2024, \apj, 976, 6

\end{thebibliography}



\end{document}